\documentclass[aip,jcp,reprint,superscriptaddress,footinbib]{revtex4-1}
\usepackage{graphicx}
\usepackage{comment}
\usepackage{amsmath}
\usepackage{wasysym}
\usepackage[usenames, dvipsnames]{color}
\usepackage{braket}
\usepackage{hyperref}
\usepackage{epstopdf}
\usepackage{subcaption}
\usepackage[symbol]{footmisc}

\begin{document}

\title{Noise and Thermodynamic Uncertainty Relation in "Underwater" Molecular Junctions}

\author{Henning Kirchberg}
\affiliation{Department of Chemistry, University of Pennsylvania, Philadelphia, Pennsylvania 19104, United States of America}
\email{khenning@sas.upenn.edu}
\author{Abraham Nitzan}
\affiliation{Department of Chemistry, University of Pennsylvania, Philadelphia, Pennsylvania 19104, United States of America}
\email{anitzan@sas.upenn.edu}

\date{\today}
\begin{abstract}
We determine the zero-frequency charge current noise in a metal-molecule-metal junction embedded in a thermal environment, e.g., a solvent, dominated by sequential charge transmission described by a classical master equation, and study its dependence of specific model parameters, i.e., the environmental reorganization energy and relaxation behavior. Interestingly, the classical current noise term has the same structure as its quantum analog which reflects a charge correlation due to the bridging molecule. We further determine the thermodynamic uncertainty relation (TUR) which defines a bound on the relationship between the average charge current, its fluctuation and the entropy production in an electrochemical junction in the Marcus regime. In a second part, we use the same methodology to calculate the current noise and the TUR for a protoype photovoltaic cell in order to predict its upper bound for the efficiency of energy conversion into useful work. 
\end{abstract}

\maketitle

\section{Introduction}
\label{intro}
Classical thermodynamics deals with general laws governing the dynamics of a macroscopic system, e.g., a heat engine or a refrigerator, which exchanges heat, energy and matter with an environment to produce useful work. As a central result of the second law, total entropy production is found to be never negative in any such process which leads to fundamental limits on the system's efficiency in transforming energy into useful work \cite{CarnotBook,cla1854}. Miniaturizing heat engines down to nanoscale have become a topic of wide interest in recent years \cite{ben2017}. An open nanosystem driven out of its thermal equilibrium can be also characterized by fluxes of energy, heat and matter between system and environment. At the nanoscale, one needs to consider the dynamics of individual microstates, where a significant progress has been achieved in the last three decades.
Stochastic thermodynamics relates the changes of the system microstates described by an ensemble of single fluctuating trajectories to macroscopic observables like heat, work and entropy production \cite{jar1997,sei2012,van2015,esp2009,sei2019}. 
Since fluctuations are ubiquitous in nanosystems away from thermal equilibrium, it is central to identify universality in their behavior. It is desirable for a nanoscale engine to have both a low entropy production rate (mostly dissipated as heat) and small noise in the measurable observables.

A remarkable result in this field is the thermodynamic uncertainty relation (TUR) \cite{bar2015,gin2016,hor2020}. The TUR is a dimensionless bound involving the averaged current $\langle J \rangle$, e.g., of particle number or energy, its variance $\langle \delta J^2 \rangle$ \cite{footnote1}, and the average entropy production $\dot{\sigma}$ at temperature $T$, that may be written in the form
\begin{align}
\label{TUR1}
\frac{\langle \delta J^2 \rangle}{2\langle J \rangle^2}\dot{\sigma}T\geq 2k_BT.
\end{align}
The TUR is a cost-noise trade-off relation between entropy production (cost) and relative fluctuation (noise). Loosely speaking, the TUR reveals that beyond a certain threshold, noise reduction can only be obtained by increasing entropy production. Therefore, the TUR can be used to obtain bounds on the entropy production of a system without a detailed knowledge about its microscopic structure \cite{pie2016}. Systems that obey this inequality satisfy the TUR which has been demonstrated in multiple realizations, e.g. biomolecular processes \cite{bar2015,pie2018,jac2020}, heat transport \cite{sar2019} and Brownian clocks \citep{bar2016} to cite some of many. Violations of the TUR inequality are predicted, e.g., in certain kinetic models with unidirectional transitions \cite{pal2021} or an underdamped dynamics in pendulum clocks \cite{pie2022}. In quantum systems, breaking TUR bounds has been noted due to coherence \cite{can2020,kal2021} or particle correlation \cite{bra2018}. In this context a multitude of theoretical studies of quantum molecular junctions have reported a violation of the TUR \cite{aga2018,pta2018,jun2019}. It has been shown that single and double quantum dot junctions in certain parameter regimes, when the charge transmission function is structured in the bias window, does not satisfy the inequality in Eq.\ \eqref{TUR1} and breaks the bound \cite{aga2018,jun2019}. However, a recent experiment of realistic molecular junctions in thermal environment reports that the TUR bound holds \cite{fri2020}. 

Molecular junctions immersed in a thermal and fluctuation environment are often described by the Marcus theory \cite{mar1956a,mar1956b,mar1986}. Charge transmission in this regime occurs by successive electron hopping between the molecule and the metal leads. The overall conductance in this case is determined by metal-molecule coupling, the solvent induced stabilization (administered by the reorganization energy), and solvent fluctuations needed to overcome the localization barrier \cite{zus1980,zus1995,zha2008}. The other extreme limit, where solvent induced relaxation is ineffective, molecular charge transmission corresponds to coherent tunneling transport that is described by the Landauer theory \cite{lan1932,bla2000}. 
A detailed investigation of statistical current fluctuations, thermodynamic properties and possible bounds of realistic electrochemical junctions based on electron hopping kinetics is still lacking. 
 
The aim of the present work is to study the current noise in electrochemical junctions described by electron hopping, the entropy production rate and the related minimal bound in thermodynamic cost-noise relation by the TUR. We investigate these properties in dependence of the junction parameters, e.g., energy level spacing, environmental relaxation and its reorganization energy and distinguish similarities and differences of the current noise and the TUR with the coherent electron tunneling case. In several recent works the TUR inequality has been used to establish bounds on thermodynamic performances based on fluctuations in experimentally easily accessible measures, e.g., the charge current \cite{pie2016,pie2018}. We use a similar analysis to examine the implication of the TUR for the performance of energy conversion to useful work in a protoype photovoltaic cell.

The paper is organized as follows: First, in Sec.\ \ref{SecI}, we determine the current noise by a Makovian master equation for a metal-molecule-metal junction in the sequential hopping limit and compare it to the charge transmission by coherent tunneling. Utilizing the Marcus charge transfer rates in Sec.\ \ref{appl}, we calculate the current noise and its dependence on the environmental reorganization energy and the molecular energy gaps.
In the following section \ref{entropy}, we determine the TUR relation for an electrochemical junction in the hopping regime and determine thermodynamic bounds in the current fluctuation-entropy production relation given the specific parameters of the model, i.e., by the energy difference of the system's state, by the reorganization energy and finite relaxation of the environment. 
Finally, in Sec.\ \ref{SecII}, we apply the concept of current fluctuation and thermodynamic cost (entropy production) to a prototype photovoltaic cell. Exploiting the TUR, we establish a bound of the performance in energy conversion of the cell into useful work by means of current fluctuations.
Sec.\ \ref{conc} concludes this paper.

\section{Current noise}
\label{SecI}
The charge motion in molecular systems coupled to bath(s) such as a solvent or electronic leads is often dominated by sequential hopping described by classical master equations. We note that such description can be shown as limiting case of the quantum dynamics derived from a microscopic Hamiltonian \cite{gal2005,gal2006} when the interaction between system and bath is small. A general master equation describes the molecule by its "microstates" $i$, e.g., different molecular energy states as used below (Eq.\ \eqref{master1}), whose probability distribution evolves according to 
\begin{align}
\label{eq1}
\frac{d}{dt}P_i(t)=\sum_j k_{ji}P_j(t),
\end{align}
where $k_{ji}$ are the transition rates from state $j$ to $i$ and $P_i(t)$ is the probability of system state $i$ at time $t$. The transition rates depend on temperature and chemical potential (determined by the bath) and satisfy the property of local detail balance reminiscent of the fact that the bath(s) always remain at thermal equilibrium. Each change of molecular system state can be related to the exchange of energy (heat) or particles (electrons) with the bath(s). Stochastic thermodynamics provides a theoretical framework to connect the dynamics of the molecular system by stochastic variables, e.g., the (continuous) change of particle number measured by a current, to the thermodynamics of the environment, e.g., its energy change \cite{van2015,sei2012,jar2011}. 
In this context, the second law of thermodynamics specifies that when a molecular system is driven out of its thermal equilibrium entropy is continuously produced \cite{PrigogineBook,tom2005,esp2010}. The entropy production rate will be discussed in Sec.\ \ref{entropy}.

We now apply such a kinetic scheme to a metal-molecule-metal conduction where the bridging molecule couples to two electronic leads to calculate the charge current and its zero-frequency noise. An applied voltage gradient drives the molecular system out of thermal equilibrium and induces a charge current. This central molecule can have at most one extra electron at any given time. We denote by states $a$ and $b$ the molecule with and without an extra electron. The transitions between these states is described by the master equation
\begin{align}
\label{master1}
\begin{pmatrix}
\dot{P}_a \\
\dot{P}_b
\end{pmatrix} 
= 
\begin{bmatrix}
-(k_{a\to b}^R+k_{a\to b}^L) & (k_{b\to a}^R+k_{b\to a}^L) \\
(k_{a\to b}^R+k_{a\to b}^L) & -(k_{b\to a}^R+k_{b\to a}^L)
\end{bmatrix}
\begin{pmatrix}
P_a \\
P_b
\end{pmatrix},
\end{align}
with the charge transfer rates $k_{a(b) \to b(a)}^K$ to the left or right leads, $K=R;L$.

We can determine the charge current and its (zero-frequency) noise by counting the number of charges $n_K$ of charge $e$ (the absolute magnitude of electron charge) interchanging with a given, say the right $K=R$, lead in a given time interval $t$. This can be seen as biased random walk with forward $k_{a\to b}^R$ and backward $k_{b\to a}^R$ rate. The number of charges $n_R$ becomes a stochastic variable of a statistical process. In the stationary state, the average particle (charge) current associated with average particle number $\langle n_R \rangle= t^{-1}\int_0^t dt' n_R(t')$ interchanging with the right lead during a time period $t$, in limit $t\to \infty$ for a stationary process, is defined as
\begin{align}
\label{current}
\langle J_R \rangle = \lim_{t\to \infty} e\langle n_R \rangle/t.
\end{align}
The zero frequency noise reads \cite{footnote1} (see Appendix\ \ref{App1})
\begin{align}
\label{noise}
\langle \delta J_R^2 \rangle &\equiv 4e^2\int_0^\infty dt \langle (J_R(t)-\langle J_R \rangle)(J_R(0)-\langle J _R\rangle)\rangle \\ \notag &= 4e^2\int_0^\infty dt \langle \delta J_R(t) \delta J_R(0) \rangle \\ \notag &= \lim_{t\to \infty} 2e^2(\langle n_R^2 \rangle - \langle n_R \rangle^2)/t,
\end{align}
where $\langle n_R^2 \rangle= t^{-1}\int_0^t dt' n_R^2(t')$ is the second moment of the particle number in time interval $t$.

We exploit the elegant expression obtained by Koza \cite{koz1999,koz2002} to determine the average charge current $\langle J_R\rangle$ and zero frequency noise $\langle \delta J_R^2 \rangle$ by the charge transfer rates. After Fourier transforming the master equation Eq.\ (\ref{master1}) by $P(w_R,w_L,t)=\sum_{n_R,n_L} \exp{(w_R n_R+w_L n_L)} P(n_L,n_R,t)$ one can formulate a modified generator, which is a $2-$dimensional square matrix $\Lambda(w_R,w_L)$ whose elements, accounting for charge exchange with the right and left lead, (see Appendix \ref{sec2} for more details) read
\begin{align}
\label{gen}
\Lambda&(w_R,w_L)
\\ \notag &= 
\begin{bmatrix}
-k_{a\to b}^R-k_{a\to b}^L & k_{b\to a}^R e^{w_R}+k_{b\to a}^Le^{-w_L} \\
k_{a\to b}^R e^{-w_R}+k_{a\to b}^Le^{w_L} & -k_{b\to a}^R-k_{b\to a}^L
\end{bmatrix}.
\end{align}
The characteristic polynomial associated with $\Lambda (w_R,w_L)$ defines a set of coefficients $C_n(w_R,w_L)$ via
\begin{align}
\det(\lambda \hat{1}-\Lambda(w_R,w_L))= \sum_n C_n(w_R,w_L)\lambda^n =0,
\end{align} 
where $\hat{1}$ represents the identity matrix. In terms of these coefficients, which are functions of the transition rates, the current and its zero frequency noise between molecule and right lead, $K=R$, can be written as \cite{koz1999} (see Appendix \ref{sec2}) 
\begin{align}
\label{chargecurr}
\langle J_R \rangle=-eC'_0/C_1,
\end{align}
and
\begin{align}
\label{zerofreq}
\langle \delta J_R^2 \rangle =-2e^2(C_0''+2C_1'\langle J \rangle+2C_2 \langle J\rangle ^2)/(C_1),
\end{align}
where $C_n\equiv C_n(w_R=0,w_L=0)$ and the primes denote derivatives with respect to $w_R$ for the right ($w_L$ for the left lead) taken at $w_R=w_L=0$. The related average charge current according to Eq.\ \eqref{chargecurr} and zero-frequency noise according to Eq.\ \eqref{zerofreq} can be determined to be equal on both sides $\langle J_R \rangle = \langle J_L \rangle = \langle J \rangle$ and $\langle \delta J_R^2 \rangle =\langle \delta J_L^2 \rangle= \langle \delta J^2 \rangle$ (see Appendix \ref{sec2}).

We find
\begin{align}
\label{eq18}
\langle J \rangle &=e\frac{k_{b\to a}^R k_{a\to b}^L-k_{a\to b}^R k_{b\to a}^L}{k_{b\to a}^R + k_{a\to b}^L+k_{a\to b}^R+ k_{b\to a}^L} \\ \notag
&=e[\langle J_{R\to L}\rangle-\langle J_{L\to R}\rangle], 
\end{align} 
where $\langle J_{R(L)\to L(R)}\rangle=e k_{b\to a}^{R(L)} k_{a\to b}^{L(R)}/C_1$ with $C_1=k_{b\to a}^R + k_{a\to b}^L+k_{a\to b}^R+ k_{b\to a}^L$, and (see Appendix \ref{sec2})
\begin{align}
\label{eq19}
 \langle \delta J^2 \rangle = \langle \delta J_1^2 \rangle - \langle \delta J_2^2 \rangle, 
\end{align}
where 
\begin{align}
\label{eq19b}
\langle \delta J_1^2 \rangle&=2e^2\frac{k_{b\to a}^R k_{a\to b}^L+k_{a\to b}^R k_{b\to a}^L}{(k_{b\to a}^R + k_{a\to b}^L+k_{a\to b}^R+ k_{b\to a}^L)}
\\ \notag & =2e[\langle J_{R\to L}\rangle+\langle J_{L\to R}\rangle] 
\\ \label{eq19c} \langle \delta J_2^2  \rangle &=2e^2\frac{2(k_{b\to a}^R k_{a\to b}^L-k_{a\to b}^R k_{b\to a}^L)^2}{(k_{b\to a}^R + k_{a\to b}^L+k_{a\to b}^R+ k_{b\to a}^L)^3} \\ \notag
&=4e[\langle J_{R\to L}\rangle-\langle J_{L\to R}\rangle]^2/C_1.
\end{align} 

It is interesting to compare these results to (a) the chemical reaction model based on a biased random walk without an intermediate step via a bridging molecule \cite{bar2015} and (b) to the corresponding quantum expression \cite{bla2000} for coherent electron transmission. In the simplest example of a nonequilibrium chemical reaction, the authors of Ref.\cite{bar2015} have proposed a biased random walk where a single step is interpreted as successful completion of a reaction. This is mathematically equivalent to a junction in which charge transfer between the left and right leads takes place directly and not via an intermediate (dot or molecule) state and the probabilities for a charge $e$ to jump from left to right and from right to left during time $\Delta t$ are given by $k^+\Delta t$ and $k^- \Delta t$, respectively. In this case it is found that the average charge current and its noise (zero frequency) are given by 
\begin{align}
\label{currentclass}
\langle J \rangle &=e(k^+ - k^-) \\ \notag
&= \langle J_{R\to L}\rangle -\langle J_{R\to L}\rangle  \\
\label{noiseclass}
\langle \delta J^2 \rangle &= 2e (k^+ + k^-) \\ \notag &=2e[ \langle J_{R\to L}\rangle +\langle J_{R\to L}\rangle],
\end{align}
where we set $\langle J_{R\to L}\rangle=ek^+$ and $\langle J_{L\to R}\rangle=ek^-$. Note that $\langle J \rangle$ (Eq.\ \eqref{currentclass}) is equivalent to Eq.\ \eqref{eq18} and $\langle \delta J^2 \rangle$ (Eq.\ \eqref{noiseclass}) to the first term $\langle \delta J_1^2 \rangle$ in Eq.\ \eqref{eq19} while no equivalent to the second term $\langle \delta J_2^2 \rangle$ in Eq.\ \eqref{eq19} can be identified. 

For the coherent electron transmission, the general quantum noise result \cite{but1990} has been recently recast in the form \cite{bra2018,jun2019} 
\begin{align}
\label{quantumnoise}
\langle \delta J^2\rangle = \langle \delta J^2\rangle_{cl}-\langle \delta J^2\rangle_{qu},
\end{align} 
where 
\begin{align}
\label{quantumnoise1}
\langle \delta J^2 \rangle_{cl} &= \frac{2e^2}{h}\int_{-\infty}^{\infty} d\epsilon\mathcal{T}(\epsilon)\{f_L(\epsilon)[1-f_R(\epsilon)] \\ \notag
&+f_R(\epsilon)[1-f_L(\epsilon)]\} \\ \notag
&= 2e\int d\epsilon (J_{L\to R}(\epsilon)+J_{R\to L}(\epsilon)),
\\ \label{quantumnoise2}
\langle \delta J^2\rangle_{qu} &= \frac{2e^2}{h}\int_{-\infty}^{\infty} d\epsilon\mathcal{T}^2(\epsilon)[f_L(\epsilon)-f_R(\epsilon)]^2 \\ \notag
& =2e\int d\epsilon (J_{L\to R}(\epsilon)-J_{R\to L}(\epsilon))^2,
\end{align}
where $J_{R\to L}=\mathcal{T}(\epsilon)f_R(\epsilon)[1-f_L(\epsilon)]$ and $J_{L\to R}=\mathcal{T}(\epsilon)f_L(\epsilon)[1-f_R(\epsilon)]$, with $\mathcal{T}(\epsilon)$ being the transmission coefficient and the Fermi functions $f_K(\epsilon)$ (Eq.\ \eqref{Fermi}) of lead $K=L;R$.

The authors of Ref.\ \cite{bra2018,jun2019} have identified the first term \eqref{quantumnoise1} in Eq.\ \eqref{quantumnoise} as the "classical" noise reminiscent of the continuous particle transfer between two classical reservoirs like in the chemical reaction model but where the additional Fermi functions accounts for the exclusion principle \cite{bra2018,jun2019}. The second term \eqref{quantumnoise2} in Eq.\ \eqref{quantumnoise} is associated with the pure "quantum" noise related to the correlated transfer of two particles \cite{bra2018,jun2019}. It is notable that the components of the quantum noise term Eq.\ \eqref{quantumnoise} can be rewritten in very similar forms to noise in the classical sequential tunneling regime Eq.\ \eqref{eq19}. Indeed, $\langle \delta J^2 \rangle_{cl}$ in Eq.\ \eqref{quantumnoise1} is equivalent to our determined term $\langle \delta J_1^2 \rangle$ (Eq.\ \eqref{eq19b}), and, $\langle \delta J^2 \rangle_{qu}$ in Eq.\ \eqref{quantumnoise2} is similar to our calculated $\langle \delta J_2^2 \rangle$ (Eq.\ \eqref{eq19c}). We see that not only the term $\langle \delta J^2 \rangle_{cl}$ (Eq.\ \eqref{quantumnoise1}) but also $\langle \delta J^2 \rangle_{qu}$ (Eq.\ \eqref{quantumnoise2}) have a clear classical analog which arises form the sequential charge transmission via the intermediate molecule. In the classical consideration of a particle current between two reservoirs without intermediate state, see Eq.\ \eqref{noiseclass}, such a term is lacking. This observation leads to the conclusion that the additional term \eqref{eq19c} and its quantum analog \eqref{quantumnoise2} reflect particle correlations. The classical correlation arises from the fact that only one electron can occupy the molecule (dot) at any given time.

More insight about these results can be obtained by considering the high bias limit, $e\Delta \Phi \gg k_B T$, where shot-noise dominates. In this case current proceeds mainly in one direction, e.g., $\langle J_{R\to L} \rangle \gg \langle J_{L\to R} \rangle$. For classical transmission without intermediate molecule, Eq.\ \eqref{noiseclass} becomes $\langle \delta J^2 \rangle = 2e \langle J \rangle$ in agreement with  the shot noise result obtained by Shottky \cite{sho1918}.
In contrast, the quantum result, Eq.\ \eqref{quantumnoise}, yields in this limit $\langle\delta J^2 \rangle=2e\langle J \rangle (1-\mathcal{T})$ for an energy independent transmission coefficient $\mathcal{T}$ \cite{les1989,but1990}. 
This reproduces the classical result in the poor transmission limit while for larger  $\mathcal{T}$ the shot noise is suppressed and vanishes when $\mathcal{T}=1$ \cite{but1990}. For an electrochemical junction with electron hopping via a bridging molecule, we obtain in the high-bias limit $\langle \delta J^2 \rangle=2e\langle J \rangle[1-2k_{b\to a}^Rk_{a\to b}^L/(k_{b\to a}^R+k_{a\to b}^L)^2]$, when one current direction dominates in Eqs.\ \eqref{eq19}-\eqref{eq19c}, e.g., $\langle J_{R\to L} \rangle \gg \langle J_{L\to R} \rangle$. Noise suppression, resulting from occupation exclusion on the intermediate state as argued above, is seen also here, albeit not the full extent obtained in the quantum expression. Notably, the case with the same rate on both sides, $k_{b\to a}^R=k_{a\to b}^L$, leads to the maximal suppression by the factor $1/2$, which is equivalent to the quantum result with transmission $\mathcal{T}=1/2$. The latter is seen in the quantum case for single-dot junctions in the weak-coupling limit and equal transmission rates to both sides \cite{aga2018}.

\section{Application to an electrochemical junction in the Marcus regime}
\label{appl}
In the following we apply the general result for zero-frequency noise to an electrochemical metal-molecule-metal junctions assuming that electron transfer kinetics is given by Marcus "hopping" rates. The rates for charging ($b\to a$) and de-charging ($a\to b$) the molecule are given by
\begin{align}
\label{rate1}
k_{a\to b}^K&=\Gamma_K\sqrt{\frac{\beta_s}{4\pi E_r}} \int_{-\infty}^{\infty} d\epsilon [1-f_K(\beta_K,\epsilon)] \\ \notag &\times\exp\bigg[-\frac{\beta_s}{4E_r}(\epsilon+E_r-\epsilon_d)^2\bigg]\\
\label{rate2}
k_{b\to a}^K&=\Gamma_K\sqrt{\frac{\beta_s}{4\pi E_r}} \int_{-\infty}^{\infty} d\epsilon f_K(\beta_K,\epsilon)\\ \notag & \times \exp\bigg[-\frac{\beta_s}{4E_r}(\epsilon_d+E_r-\epsilon)^2\bigg],
\end{align}
where $\Gamma_K=2\pi/\hbar |V_K|^2 \rho_K$ are the golden rule rates for electrons moving between a discrete molecular level and a continuum of single electronic states of the left and right electrodes $K=L,R$ with the density of states $\rho_K$ which we assume independent of the energy $\epsilon$. The level-lead coupling is $V_K$. $\beta_s=(k_B T_s)^{-1}$ is the thermal energy of the environment, $E_r$ is the reorganization energy associated with relaxation of the nuclear environment following electron hopping events and $\epsilon_d$ is the energy difference between the equilibrium states of the charged and uncharged molecule. The Fermi distribution of the electronic states in the metal leads $K$ at bias potential $\Delta \Phi_K$ reads
\begin{align}
\label{Fermi}
f_K(\beta_K,\epsilon)=\frac{1}{\exp\big[\beta_K(\epsilon-\mu_K-e\Delta \Phi_K)\big]+1},
\end{align} 
where the thermal energy of the electronic reservoirs is $\beta_K=(k_B T_K)^{-1}$ and their chemical potential $\mu_K$. In what follows we assume that $\mu_L=\mu_R$ and set our energy scale so that $\mu_R=\mu_L=0$. Therefore, the difference between the left and right lead stems from the applied bias. Furthermore, we express the applied bias in terms of $\Delta \Phi_L$ and $\Delta \Phi_R$ of the left and right leads, taking $\Delta \Phi_K=0$ at the molecule while the state difference $\epsilon_d$ is positive and lays above the Fermi energies of the leads at zero bias.

\begin{figure}[h!!!!!]
\centering
\includegraphics[width=\linewidth]{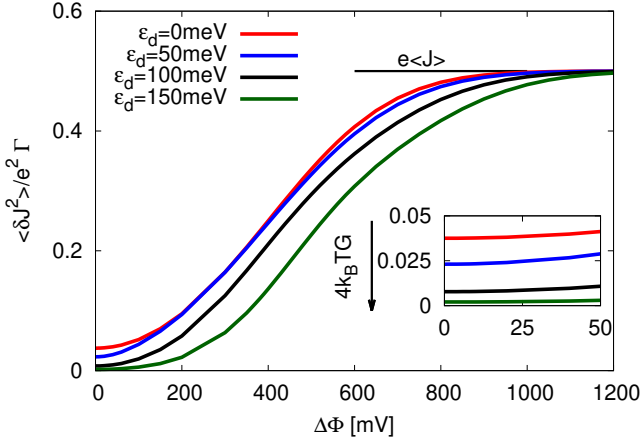}
\caption{\label{fig1a} Zero-frequency noise $\langle \delta J^2 \rangle$ plotted against different applied bias potential $\Delta \Phi$ for several values of the energy difference $\epsilon_d$ between the charged and uncharged state. Inset: $\langle \delta J^2 \rangle$ for a smaller voltage regime. The temperature is $T=300$K and $k_BT\simeq 26$meV. The reorganization energy $E_r=200$meV.}
\end{figure}

\begin{figure}[h!!!!!]
\centering
\includegraphics[width=\linewidth]{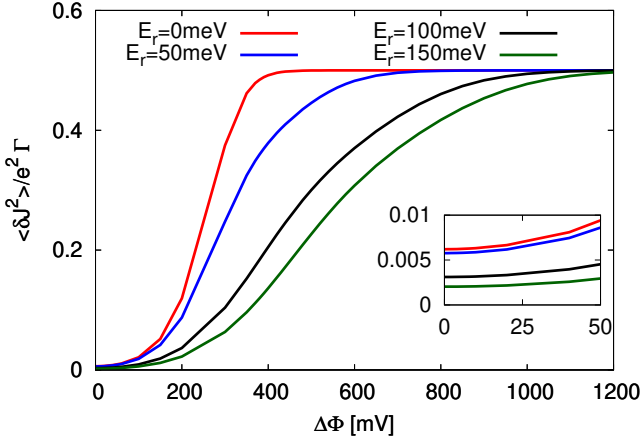}
\caption{\label{fig1aa} Zero-frequency noise $\langle \delta J^2 \rangle$ plotted against different applied bias potential $\Delta \Phi$ for several reorganization energies $E_r$. Inset: $\langle \delta J^2 \rangle$ vs. $\Delta \Phi$ for a smaller voltage regime about $\Delta \Phi=0$. The difference of equilibrium energy between the charged and uncharged state is $\epsilon_d=150$meV. The temperature is $T=300$K and $k_BT\simeq 26$meV. Note that for $E_r=0$ we set the rate \eqref{rate1} to $k_{a\to b}^K=\Gamma_K [1-f_K(\beta_K,\epsilon_d)]$ and the rate \eqref{rate2} to $k_{b\to a}^K=\Gamma_K f_K(\beta_K,\epsilon_d)$ where $\Gamma_K\equiv \Gamma$.}
\end{figure}

\begin{figure}[h!!!!!]
\centering
\includegraphics[width=\linewidth]{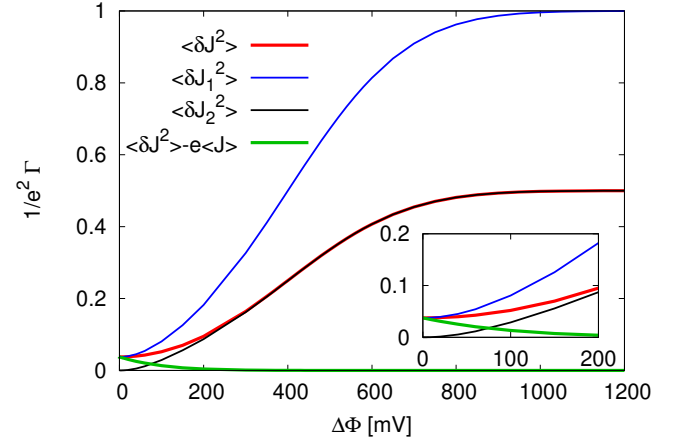}
\caption{\label{fig1b} Total zero-frequency noise, its contributions $\langle \delta J_1 ^2 \rangle$ and $\langle \delta J_2 ^2 \rangle$ (see Eq.\ \eqref{eq19}) as well as the "thermal" part $\langle \delta J ^2 \rangle-e\langle J \rangle$ plotted against bias potential for $\epsilon_d=0$. Inset: Smaller voltage regime. The temperature is $T=300$K and $k_BT\simeq 26$meV. The reorganization energy $E_r=200$meV.}
\end{figure}

Figs.\ \ref{fig1a}, \ref{fig1aa}  and \ref{fig1b} show the steady state zero-frequency noise, computed from Eq.\ \eqref{eq19}, for the Marcus model using the rates of Eqs.\ \eqref{rate1} and \eqref{rate2}. Here we have taken an uniform temperature of leads and solvent, $T_S=T_K=T$ and equal rates, $\Gamma_L=\Gamma_R=\Gamma$, to the left and right electrode. The voltage bias is applied symmetrically $\Delta \Phi_R=-\Delta \Phi_L=\Delta \Phi/2$ while keeping the energy of the molecular orbital $\epsilon_d$ fixed (at a value that may be changed independently, reflecting the effect of a gate potential).
The following observations are noteworthy:

(a) In the limit of zero bias $\Delta \Phi = 0$, the system is in equilibrium and the net charge current vanishes. The (charge) current noise in Fig.\ \ref{fig1a} is the Johnson-Nyquist thermal noise \cite{joh1927,joh1928,nyq1928} $\langle \delta J^2\rangle =4k_BTG$ where $G$ is the conductance\cite{mig2012}
\begin{align}
\label{conduc}
G&=\lim_{\Delta \Phi \to 0} d \langle J \rangle /d \Delta \Phi \\ \notag
&=\frac{e^2}{k_BT}\frac{k_{b\to a}^R(0) k_{a\to b}^L(0)}{k_{b\to a}^R(0) + k_{a\to b}^L(0)+k_{a\to b}^R(0)+ k_{b\to a}^L(0)}.
\end{align}

(b) The zero bias noise reflects the dependence of the conductance $G$ on junction parameters. Specifically it is maximal for $\epsilon_d=0$ (Fig.\ \ref{fig1a}) and decreases with increasing reorganization energy $E_r$ (Fig.\ \ref{fig1aa}) as implied by Eq.\ \eqref{conduc} and investigated in Ref.\ \cite{mig2012}. Increasing $\epsilon_d$ or $E_r$ shifts the transmission window (given by the Gaussian term in the rates of Eqs.\ \eqref{rate1} and \eqref{rate2}) out of the range to find a metal state occupied or empty (given by the Fermi functions in Eqs.\ \eqref{rate1} and \eqref{rate2}), such that at least one rate for molecular state occupation ($k_{b\to a}^K$) or deoccupation ($k_{a\to b}^K$) decreases and so their product in Eq.\ \eqref{conduc} which reduces the conductance $G$.

(c) As $\Delta \Phi$ increases the noise increases above the (thermal) Johnson-Nyquist value and saturates when $e\Delta \Phi \gg k_BT$ (see Figs.\ \ref{fig1a} and \ref{fig1aa}) at the value that is determined by the saturation current $\langle J \rangle = e(\Gamma_R \Gamma_L)/(\Gamma_R + \Gamma_L)=e\Gamma/2$ for $\Gamma_R=\Gamma_L=\Gamma$. In this limit Eq.\ \eqref{eq19} reduces to $\langle \delta J^2\rangle=e\langle J \rangle$ for the single intermediate level model. 
Note that the Schottky result $2e\langle J \rangle$ is suppressed by a factor $2$ as explained at the end of Sec.\ \ref{SecI}. 

(d) Fig.\ \ref{fig1b} portrays the two components $\langle \delta J^2_1 \rangle$ (Eq.\ \eqref{eq19b}) and $\langle \delta J^2_2 \rangle$ (Eq.\ \eqref{eq19c}), as well as the total noise $\langle \delta J^2\rangle = \langle \delta J_1^2 \rangle - \langle \delta J_2^2 \rangle$, Eq.\ \eqref{eq19}. For the small voltage limit $e\Delta \Phi \ll k_BT$, the total noise can is captured by the first component, $\langle \delta J^2 \rangle\to \langle \delta J^2_{1} \rangle$, while $\langle \delta J^2_{2} \rangle$ does not contribute. In this regime, $\langle \delta J^2 \rangle = \langle \delta J^2_{1} \rangle$ is associated to the thermal noise. Away from thermal equilibrium, for finite bias potential, the second term $\langle \delta J_2^2 \rangle$, associated with charge correlation, must be also considered which leads to a reduction of the total noise by $\langle \delta J^2 \rangle= \langle \delta J_1^2 \rangle-\langle \delta J_2^2 \rangle=2e\langle J \rangle[1-2k_{b\to a}^Rk_{a\to b}^L/(k_{b\to a}^R+k_{a\to b}^L)^2]$. We see a maximal reduction of a factor not smaller than $2$ for equal transfer rates, $k_{b\to a}^R=k_{a\to b}^L$ (see discussion end of Sec.\ \ref{SecI}). However, for highly asymmetric rates, the term $\langle \delta J_2^2 \rangle$ term becomes very small also for $\Delta \Phi > 0$ so that $\langle \delta J_1^2 \rangle$ dominates.

(e) Fig.\ \ref{fig1b} also shows the difference $\langle \delta J^2 \rangle-e\langle J \rangle$ (see green line Fig.\ \ref{fig1b}) between the total noise and its higher bias (shot noise) limit. The difference $\langle \delta J^2 \rangle-e\langle J \rangle$ might be associated to the thermal noise for finite bias potential. For zero temperature this term vanishes and the current and its noise would show a sharp onset when potential energy equals the molecular level $e\Delta \Phi = 2\epsilon_d$. For finite temperature, as shown in Fig.\ \ref{fig1b}, $\langle \delta J^2 \rangle-e\langle J \rangle$ smoothly goes to zero in the voltage regime $e\Delta \Phi \sim 2k_BT$.

Finally, we note that the temperature dependence of the current noise can be used as a temperature monitoring device. Indeed, Spietz et al. \cite{spi2003} present an electronic thermometer where they read off the temperature from equilibrium Johnson-Nyquist noise at zero voltage. 

\section{Entropy production and thermodynamic uncertainty relation}
\label{entropy}
For a system whose state dynamics is described by the master equation of Eq.\ (\ref{eq1}), the entropy production rate $\dot{\sigma}$ can be derived from the Boltzmann-Gibbs expression \cite{bol1877,GibbsBook} for the total system entropy $S(t)=-k_B\sum_iP_i(t)\ln P_i(t)$ (see Appendix \ref{sec3}) \cite{sna1976}
\begin{align}
\label{eq2}
\dot{\sigma}(t)=\frac{k_B}{2}\sum_{ij}[k_{ji}P_j(t)-k_{ij}P_i(t)]\ln{\frac{k_{ji}P_j(t)}{k_{ij}P_i(t)}}.
\end{align}
The expression (\ref{eq2}) meets two important properties: (i) It is nonegative because each term in the summation is of the form $(x-y)\ln(x/y)>0$ and (ii) vanishes for thermal equilibrium, when microscopic reversibility or detailed balance condition, $k_{ji}P_j=k_{ij}P_i$, is obeyed and no entropy is produced. 

At (nonequilibrium) steady state with the respective $P_{i}$ stationary probability distribution the entropy production reduces to the expression \cite{tom2005,esp2010} (see Appendix \ref{sec3})
\begin{align}
\label{eq3}
\dot{\sigma}&=\frac{k_B}{2}\sum_{ij} [k_{ji}P_j-k_{ij}P_i]\ln{\frac{k_{ji}}{k_{ij}}} \\ \notag
&=\frac{k_B}{2}\sum_{ij} J_{ij}A_{ij},
\end{align}
where $J_{ij}=k_{ji}P_j-k_{ij}P_i=-J_{ji}$ is the current between states $i$ and $j$ which is driven by the corresponding force or affinity $A_{ij}=\ln{k_{ji}/k_{ij}}$.

With the entropy production rate $\dot{\sigma}$ at steady state define the product
\begin{align}
\label{TUR2}
Q \equiv \dot{\sigma} T \langle \delta J^2 \rangle/2 \langle J\rangle ^2, 
\end{align}
where $\langle J \rangle$, Eq.\ \eqref{chargecurr}, is the average charge current and $\langle\delta J^2\rangle$, Eq.\ \eqref{zerofreq}, the related zero-frequency noise.
The inequality $Q\geq 2k_BT $ (Eq.\ \eqref{TUR1}) is the thermodynamic uncertainty relation (TUR). This inequality has been shown to be satisfied for a large variety of systems \cite{bar2015,dec2018,hor2020}. $Q$ will be henceforth referred as the TUR product.

By identifying the entropy production rate as Joule heating, $\dot{\sigma}=\Delta \Phi \langle J \rangle/T$, $Q$ in our metal-molecule-metal junction may be written in other physically appealing forms. In terms of $g\equiv \langle J\rangle /\Delta \Phi$, Eq. \eqref{TUR2} can be recast as
\begin{align}
\label{TUR3}
Q= \frac{\langle \delta J^2 \rangle}{2g},
\end{align}
while in terms of the Fano factor, $F=\langle \delta J^2 \rangle/\langle J \rangle$, $Q$ takes the form
\begin{align}
\label{TUR4}
Q= \Delta \Phi \frac{F}{2},
\end{align}
which will be used later.

Using Eqs.\ \eqref{rate1} and \eqref{rate2}, the TUR product in Eq.\ \eqref{TUR2} yields
\begin{align}
\label{eq20}
Q &=Q_{1}-Q_{2},
\end{align}
with
\begin{align}
\label{eq20a} Q_1&=e\Delta \Phi \frac{k_{b\to a}^R k_{a\to b}^L+k_{a\to b}^R k_{b\to a}^L}{k_{b\to a}^R k_{a\to b}^L-k_{a\to b}^R k_{b\to a}^L}
\\ \label{eq20aa} &=e\Delta \Phi \coth\big[\beta e \Delta \Phi/2\big],
\end{align}
and
\begin{align}
 \label{eq20b} Q_2&=2e\Delta \Phi\frac{k_{b\to a}^R k_{a\to b}^L-k_{a\to b}^R k_{b\to a}^L}{(k_{b\to a}^R + k_{a\to b}^L+k_{a\to b}^R+ k_{b\to a}^L)^2}.
\end{align}

The form for $Q_1$ in Eq.\ \eqref{eq20aa} is obtained by using the detailed balance relation of the forward and backward hopping rates (see Eq.\ \eqref{1} Appendix \ref{sec4}). $Q_1$ is equal to the total TUR product obtained for a rate process described by a simple one dimensional biased random walk that underlines hopping without an intermediate reaction center \cite{bar2015}, see Sec.\ \ref{SecI}. It is easy to show form Eq.\ \eqref{eq20aa} that $Q_{1} \geq 2k_BT$ as expected\cite{bar2015}. Also, it is easily realized that the term $Q_2\geq 0$, since $k_{b\to a}^R k_{a\to b}^L>k_{b\to a}^R k_{a\to b}^L$ for $\Delta \Phi>0$ and $k_{b\to a}^R k_{a\to b}^L<k_{b\to a}^R k_{a\to b}^L$ for $\Delta \Phi<0$ in Eq.\ \eqref{eq20b}, which holds the positivity of the TUR product. For the particular case where (a) $\Gamma_L=\Gamma_R$, (b) $T_K=T_S=T$ and (c) $\epsilon_d=0$ (i.e. $\epsilon_d$ is equal to the unbiased left and right Fermi energies), it follows from Eqs.\ \eqref{1}, \eqref{2b} and \eqref{3} in Appendix \ref{sec4}, for the relation between the transfer rates, that $Q_{2}$ of Eq.\ \eqref{eq20b} can be written in the from 
\begin{align}
\label{6}
Q_2&=e\Delta \Phi\frac{1}{4}\bigg[\tanh\big[\beta e \Delta\Phi/4\big]-\tanh\big[-\beta e \Delta\Phi/4\big]\bigg]\\ \notag
&=\frac{e\Delta \Phi}{2}\tanh\big[\beta e \Delta\Phi/4\big].
\end{align}
Since $Q_2\geq 0$, it can potentially lead to violation of the inequality formulated in Eq.\ \eqref{TUR1} for $Q=Q_1-Q_2$, however we show in the following discussion that this is not the case in our classical hopping model.

\begin{figure}[h!!!!!]
\centering
\includegraphics[width=\linewidth]{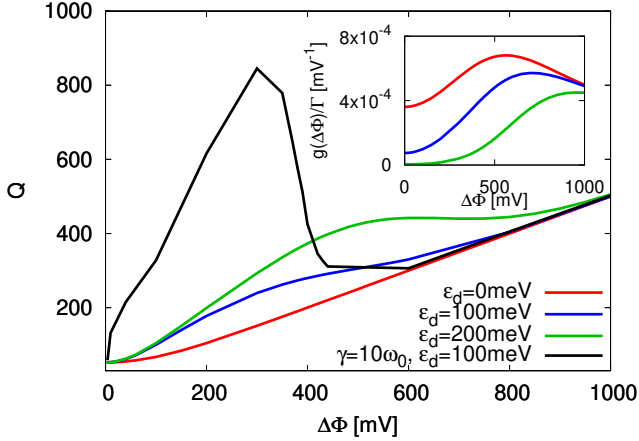}
\caption{\label{fig3} The TUR product $Q$ displayed against the applied bias potential $\Delta \Phi$ for several values of the energy difference $\epsilon_d$ between the charged and uncharged state. The black line represents $Q$ calculated within the high-friction model (see text). Inset: $g=\langle J \rangle/\Delta \Phi$ for the same choices of $\epsilon_d$. The temperature is $T=300$K and $k_BT\simeq 26$meV. The reorganization energy $E_r=200$meV.}
\end{figure}
 
Note that the result in Eq.\ (\ref{eq20}) resembles the functional form obtained by Liu and Segal \cite{jun2019} for the quantum resonance model that can be also written as $Q^{qu}=Q_1^{qu}-Q_2^{qu}$, where $Q_1^{qu}=Q_1$ (Eq.\ \eqref{eq20aa}), while for certain resonance transmission models, i.e. a resonant single-dot or a serial double-dot junction \cite{aga2018,jun2019},
\begin{align}
\label{7}
Q_2^{qu}=e\Delta \Phi\frac{\Theta}{2}\bigg[\tanh\big[\beta e \Delta\Phi/4\big]-\tanh\big[-\beta e \Delta\Phi/4\big]\bigg],
\end{align}
where $\Theta\equiv \int d\epsilon T(\epsilon)^2/\int d\epsilon T(\epsilon)$. In order to check a possible violation of TUR, the authors\cite{jun2019} have investigated the condition $Q^{qu}-2k_BT\le 0$ and found $\Theta>2/3$ for violating TUR. The result in the sequential hopping regime (Eq.\ \eqref{eq20b}) is the same except the $\Theta$ is replaced by $1/2$ for the symmetrical hopping regime ($\Gamma_R=\Gamma_L$). This case yields the maximum value of $Q_2$ and it therefore follows that the TUR bound of $2k_BT$ can not be violated in the classical hopping regime.

\begin{figure}[h!!!!!]
\centering
\includegraphics[width=\linewidth]{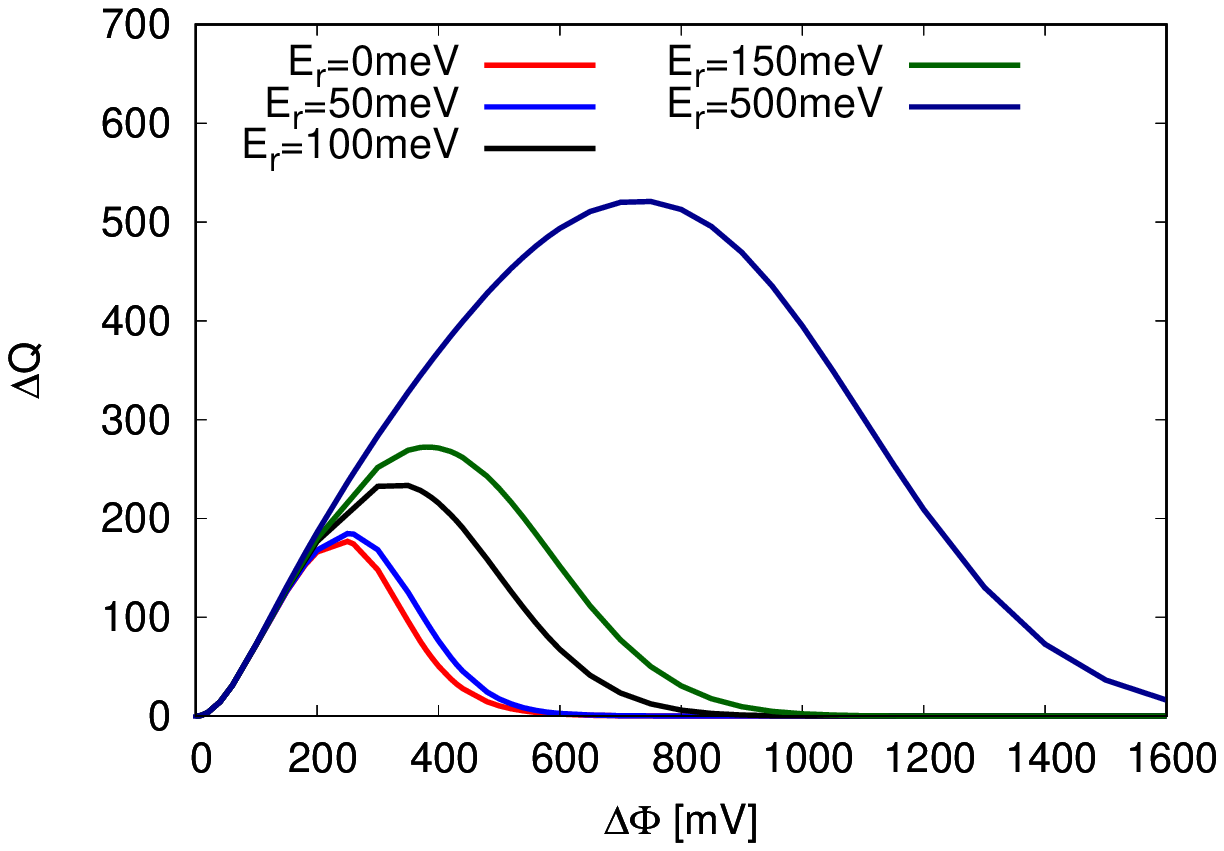}
\caption{\label{fig3a} $\Delta Q =Q-e\Delta \Phi[\coth[\beta e\Delta \Phi/2]-\tanh[\beta e\Delta \Phi/4]/2]$ against different applied bias potential $\Delta \Phi$ for several reorganization energies $E_r$. The temperature is $T=300$K and $k_BT\simeq 26$meV. The difference between equilibrium energies between the occupied and unoccupied states is $\epsilon_d=150$meV. Note that for $E_r=0$ we set the rate \eqref{rate1} to $k_{a\to b}^K=\Gamma_K [1-f_K(\beta_K,\epsilon_d)]$ and the rate \eqref{rate2} to $k_{b\to a}^K=\Gamma_K f_K(\beta_K,\epsilon_d)$ where $\Gamma_K\equiv \Gamma$.}
\end{figure}

\begin{figure}[h!!!!!]
\centering
\includegraphics[width=\linewidth]{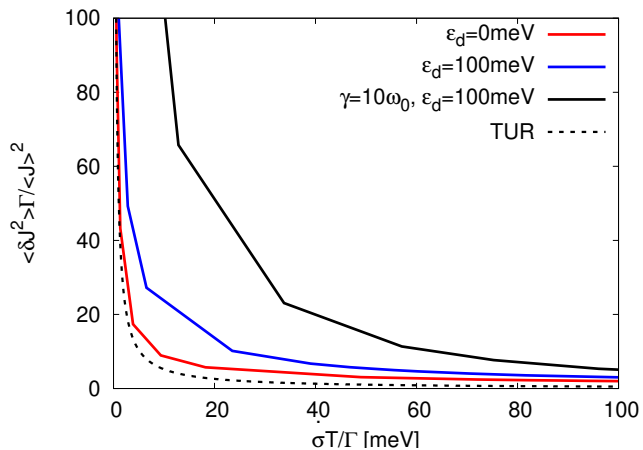}
\caption{\label{fig3b} Zero-frequency noise to current ratio plotted against entropy production rate for several values of the energy difference $\epsilon_d$ between the charged and uncharged state. The black dotted line represents the minimal TUR bound $\langle \delta J ^2 \rangle/\langle J \rangle ^2=4k_BT/\dot{\sigma}T$. The black solid line represents $Q$ calculated within the high-friction model (see text). The temperature is $T=300$K and $k_BT\simeq 26$meV. The reorganization energy $E_r=200$meV.}
\end{figure}

Examples of the behavior of the TUR product, Eq.\ \eqref{eq20}, in our model molecular junction by sequential hopping are shown in Figs.\ \ref{fig3}-\ref{fig3b}, where we have used the expressions \eqref{rate1} and \eqref{rate2} (Marcus expression) for the kinetic rates. We assume again a uniform temperature $T_S=T_K=T$ as well as equal rates $\Gamma_L=\Gamma_R=\Gamma$ and apply the voltage symmetrically $\Delta \Phi_R=-\Delta \Phi_L=\Delta \Phi/2$. Fig.\ \ref{fig3} portrays the bias voltage dependence of the TUR product $Q$ plotted for different values of $\epsilon_d$. For a vanishing voltage $\Delta \Phi \to 0$ (thermal equilibrium) all curves are bounded by $2k_BT$. With increasing voltage $Q$ increases as both the entropy production $\dot{\sigma}=\Delta \Phi \langle J\rangle/T$ and the relative fluctuation $\langle \delta J^2 \rangle/\langle J \rangle^2$ increase. The result for $\epsilon_d=0$ (red curve in Fig.\ \ref{fig3}) was obtained both from the general expressions Eqs.\ \eqref{eq20a} and \eqref{eq20b} and from the analytic results in Eqs.\ \eqref{eq20aa} and \eqref{6} which is independent of the reorganization energy $E_r$ of the solvent. We prove that $Q_2^{\epsilon_d=0} \geq Q_2^{\epsilon_d\neq 0}$ (see Appendix \ref{sec5}) such that the TUR product $Q=Q_1-Q_2$, Eq.\ \eqref{eq20}, is minimal for $\epsilon_d=0$ for a junction of equal temperature in leads and solvent. This intriguing result yields a minimal bound on the product of entropy production $\dot{\sigma}$ and relative fluctuation $\langle \delta J^2 \rangle/\langle J \rangle^2$. 

From the inset to Fig.\ \ref{fig3}, it appears that $Q$ is mostly dominated by the strong dependence of $g=\langle J \rangle/\Delta \Phi$ on $\Delta \Phi$ (see $Q$ written in the form of Eq.\ \eqref{TUR3}). This in turn implies that $Q$ is smaller at $\epsilon_d=0$
than at finite $\epsilon_d$ and also (since $g$ is maximal at $\epsilon_d=0$) that $Q$ increases faster with $\Delta \Phi$ when $\epsilon_d=0$ than when $\epsilon_d$ is larger. Note that the dependence of $\langle \delta J^2 \rangle$ on $Q$ (Eq.\ \eqref{TUR3}) is relatively weak since the dependence of $\langle \delta J^2 \rangle$ on $\Delta \Phi$ and different values of $\epsilon_d$ is only visible in the thermal noise regime (see $ e\Delta \Phi \ll k_BT$ in Fig.\ \ref{fig1a}).
 
The use of Marcus rates in the sequential electron transfer process considered above relies on the assumption of fast relaxation of the solvent in between each electron hop. In our recent works \cite{kir2020,kir2022} we account for finite solvent relaxation where the transfer rates \eqref{rate1} and \eqref{rate2} becomes explicitly time-dependent (see details in Ref. \cite{kir2020,kir2022}). The black line in Fig.\ \ref{fig3} portrays the TUR result for finite solvent relaxation described by an overdamped solvent coordinate (here damping $\gamma= 10\omega_0$, see details in Ref. \cite{kir2020}). We have shown that the Fano factor $F=\langle \delta J^2 \rangle/\langle J\rangle$ is larger for slower solvent relaxation \cite{kir2020}. Since $Q\propto F$ (Eq.\ \eqref{TUR4}), $Q$ grows even faster for increasing bias potential $\Delta \Phi$ in this case than in the Marcus regime (compare black and red lines in Fig.\ \ref{fig3}). The increased Fano factor ($F>1$) and the related increase in (relative) current fluctuation for finite solvent relaxation can be understood as correlated (non-Poissonian) electron transfer.

Eqs.\ \eqref{eq20aa} and \eqref{6} have led to the observation that \\ $Q=e\Delta\Phi[\coth[\beta e \Delta \Phi/2]-\tanh[\beta e \Delta \Phi/4]/2]$ for a junction with uniform temperature $T$, $\Gamma_R=\Gamma_L=\Gamma$ and $\epsilon_d=0$, regardless of the value of $E_r$. Next, we examine the dependence of $Q$ on $E_r$ for a similar junction except with $\epsilon_d\neq 0$. Fig.\ \ref{fig3a} portrays the difference $\Delta Q = Q-e\Delta\Phi \big[\coth[\beta e \Delta \Phi/2]-\tanh[\beta e \Delta \Phi/4]/2]\big]$. $\Delta Q$ grows with higher values of $E_r$. This can be understood by the smaller $g(\Delta \Phi)$ for increased reorganization energy $E_r$ which dominates and increases $Q=\langle \delta J^2\rangle/2g$ (Eq.\ \eqref{TUR3}). This lower rise of $g(\Delta \Phi)$ (see Fig.\ \ref{figConduct} in Appendix \ref{sec6}) can be understood as part of the applied bias potential is used to overcome the reorganization energy \cite{mig2012}.

A recent experimental study \cite{fri2020} of a realistic single atomic junction confirms that the TUR product has a minimal bound of $2k_BT$ in presence of a thermal environment. For a prototype underwater junction at uniform temperature governed by the Marcus transfer kinetics our analytic result of $Q_1$ in Eq.\ \eqref{eq20aa} together with $Q_2$ in Eq.\ \eqref{6} for $\epsilon_d=0$ serves as minimal cost-fluctuation bound for different applied bias potentials.

Another perspective on the TUR is to consider the relative fluctuation $\langle \delta J^2 \rangle/\langle J \rangle^2$ against the consumed power given by $\dot{\sigma} T=\Delta\Phi \langle J \rangle$, see Fig.\ \ref{fig3b}. Both, relative fluctuation and consumed power, depend on the applied bias potential and on the transfer model specific parameters which leads to the following observations:
For the same (possibly small) relative fluctuations the minimal power consumed increases with $\epsilon_d$ and is bound from below by the (red) curve of $\epsilon_d=0$ in Fig. \ref{fig3b}. This can be understood: More energy is consumed per time to measure the same $\langle \delta J^2 \rangle/\langle J \rangle^2$ since the equilibrium energy difference between the unoccupied and occupied molecular state needs to be overcome for each charge transferred. 
Even more energy needs to be consumed to remain at a given small relative fluctuation (current noise to current), see solid black line in Fig.\ \ref{fig3b}, when the solvent relaxes slowly in between the sequential charge hops (beyond Marcus theory) and the charge transfer dynamics becomes strongly correlated (see discussion above). 
All fluctuation-cost curves a bound by the curve $\langle \delta J ^2 \rangle/\langle J \rangle ^2=4k_BT/\dot{\sigma}T$ (black dotted line in Fig.\ \ref{fig3b}) which corresponds to $Q=2k_BT$ in Eq.\ \eqref{eq20}.

In summary, we have seen that the voltage dependence of the TUR product $Q$ reflects mainly the dependence of $g(\Delta \Phi)=\langle J \rangle /\Delta \Phi$ on the bias potential for the specific junction parameters ($E_r$ and $\epsilon_d$). The zero-frequency noise $\langle \delta J^2 \rangle$ is less sensitive (in relative values) to these parameters and show a sensitivity only for the bias potential in the thermal noise regime ($e\Delta \Phi \ll k_BT$), see discussion in Sec.\ \ref{SecI}. As it is often the case, we find that reducing relative fluctuation is most readily achieved by increasing the signal $\langle J \rangle$ while our control of the noise $\langle \delta J^2 \rangle$ is relatively limited. As noted above, it is only near $\Delta \Phi=0$ (where thermal noise dominates) that the excess noise $\langle \delta J^2 \rangle_{\Delta \Phi}-\langle \delta J^2 \rangle_{\Delta \Phi=0}$ is sensitive to system parameters. In particular, it is easy to show by the same arguments as for the minimal bound of the term $Q$ that the change $d (\langle \delta J^2 \rangle/\langle J \rangle^2)/d \Delta \Phi_{|\Delta\Phi=0}$ (which is positive) is smallest for $\epsilon_d=0$.

\section{Application: prototype photovoltaic cell}
\label{SecII}
We next calculate the stationary charge current, its fluctuation and the resulting TUR product of the prototype photovoltaic cell studied earlier by Rutten, Esposito and Cleuren \cite{rut2009}. 
\begin{figure}[h!!!!!]
\centering
\includegraphics[width=\linewidth]{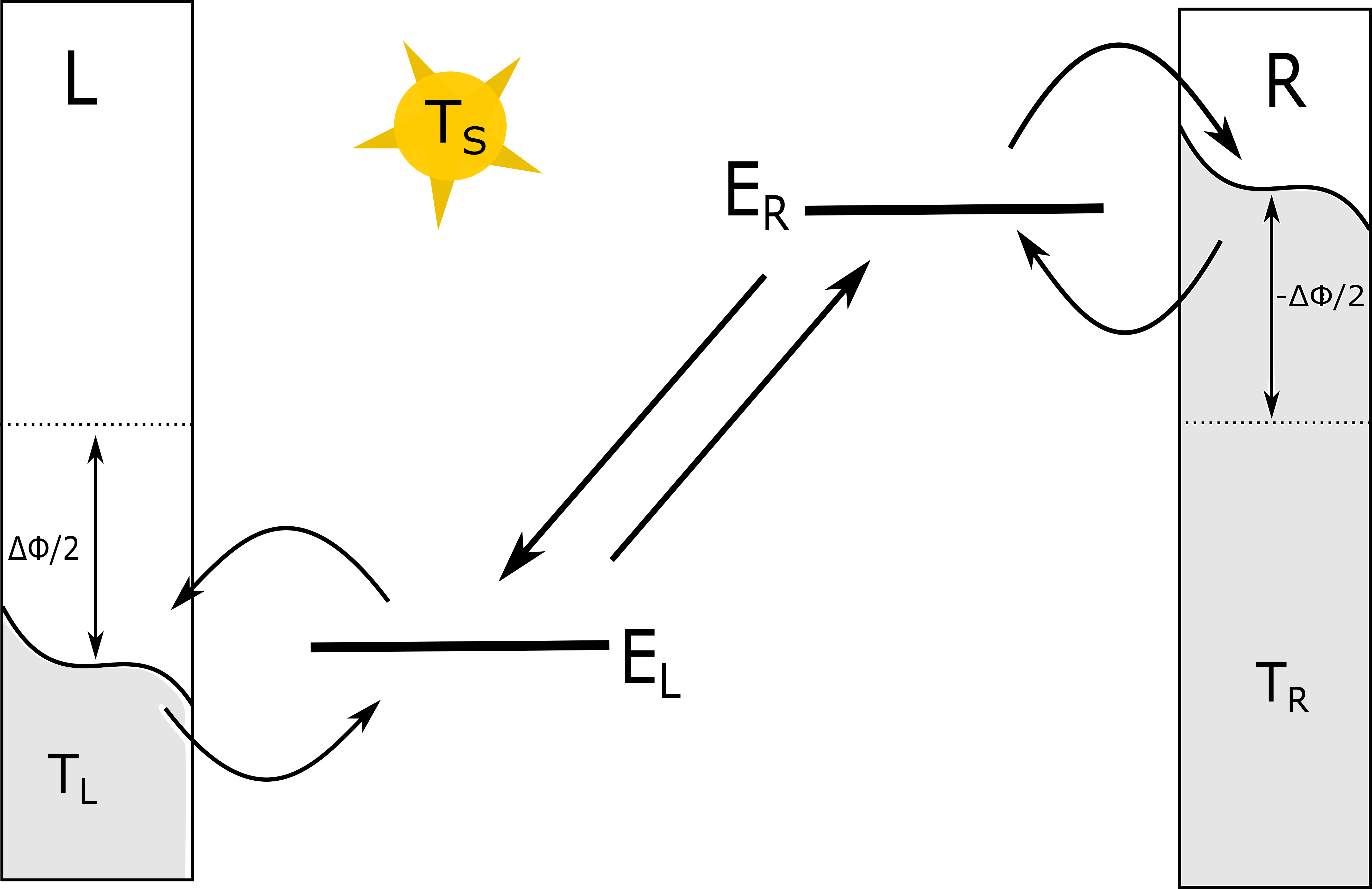}
\caption{\label{Model} Model of the photovoltaic cell.}
\end{figure}
This device model, see Fig.\ \ref{Model}, is composed of two single particle levels of energy $E_L$ and $E_R (>E_L)$ which define the bandgap $\Delta E=E_R-E_L$. The Coulomb repulsion is assumed to restrict the possible system states to $0,L$ and $R$ with occupation zero or one electron on level $E_L$ or $E_R$, respectively. The leads are at the same temperature $T_L=T_R=T$ where we assume that level $E_K$ can exchange electron only with lead $K$ ($K=L,R$). As before we define our origin of single electron energy by setting the Fermi energy of the unbiased lead to $E_F=0$. Under bias the leads then have different chemical potential $e\Delta \Phi_K=\pm e \Delta \Phi/2$ (positive when the chemical potential is higher on the left).
Electron transition between $E_R$ and $E_L$ are induced by the incoming "sun" radiation which is assumed to be at resonant energy $\Delta E = h \nu$. The dynamics of this cell is described by using the master equation for the probabilities $P_j(j=0,L,R)$ to be in the corresponding states which reads
\begin{align}
\label{masterphoto}
\begin{pmatrix} 
		\dot{P}_0(t) \\ 
		\dot{P}_L(t) \\ 
		\dot{P}_R(t)
	\end{pmatrix} 
	=&
\begin{bmatrix} 
		-(k_{0L}+k_{0R}) & k_{L0} & k_{R0}  \\ 
		k_{0L} & -(k_{L0}+k_{LR}) & k_{RL} \\ 
		k_{0R} & k_{LR} & -(k_{R0}+k_{RL})
	\end{bmatrix} \\ \notag &\times \begin{pmatrix} 
		P_0(t) \\ 
		P_L(t) \\ 
		P_R(t)
	\end{pmatrix},
\end{align}
where $k_{ji}$ denotes the transition from state $j$ to $i$. 
The rates describing the exchange of electrons with the leads are given by
\begin{align}
\label{ratesleads}
k_{0L}&=\Gamma_L f(x_L); \hspace*{0.5cm} k_{L0}=\Gamma_L (1-f(x_L)) \\
\label{ratesleads2}
k_{0R}&=\Gamma_R f(x_R); \hspace*{0.5cm}
k_{R0}=\Gamma_R (1-f(x_R)),
\end{align}
where $f(x)=[\exp(x)+1]^{-1}$ is the Fermi distribution where $x_L = \beta(E_L-e\Delta \Phi/2)$ and $x_R=\beta(E_R+e\Delta \Phi/2)$ with the inverse temperature $\beta^{-1}=k_BT$. We assume again equal transfer rates $\Gamma_L=\Gamma_R=\Gamma$ to both leads.
The rates describing the transition between energy level $E_L$ and $E_R$ due to sun photons are given by
\begin{align}
\label{ratessun}
k_{LR}&=\Gamma_S n(x_S) \\
\label{ratessun2}
k_{RL}&=\Gamma_S [1+n(x_S)],
\end{align}
where $n(x)=[\exp(x)-1]^{-1}$ is the Bose-Einstein distribution with $x_S=\beta_S \Delta E$ with the inverse temperature $\beta_S^{-1}=k_BT_S$. Note that the ratio between forward and backward transition rates in Eqs.\ \eqref{ratesleads} and \eqref{ratesleads2} as well as \eqref{ratessun} and \eqref{ratessun2} satisfy the detailed balance condition.  For simplicity, we neglect nonradiative transitions between the molecular levels $E_R$ and $E_L$.
We utilize again the formalism of Ref. \cite{koz1999} and use the modified generator of the master equation \eqref{masterphoto} to determine the average current and its zero-frequency noise. This procedure (see Appendix \ref{sec7}) leads to
\begin{align}
\label{photocurr}
\langle J \rangle = -e\frac{k_{L0}k_{RL}k_{0R}-k_{0L}k_{LR}k_{R0}}{C_1},
\end{align}
and 
\begin{align}
\label{photonoise}
\langle \delta J^2 \rangle &= 2e^2 \frac{(k_{L0}k_{RL}k_{0R}+k_{0L}k_{LR}k_{R0})}{C_1} \\ \notag
&-2e^2 \frac{(k_{0L}+k_{0R}+k_{L0}+k_{LR}+k_{R0}+k_{RL})}{C_1^3}\\ \notag &\times 2(k_{L0}k_{RL}k_{0R}-k_{0L}k_{LR}k_{R0})^2 ,
\end{align}
where $C_1=k_{R0}(k_{L0}+k_{LR}+k_{0L})+k_{RL}(k_{0L}+k_{0R}+k_{L0})+k_{0R}k_{L0}+k_{LR}(k_{0L}+k_{0R})$.

The affinity $A$ for the device reads
\begin{align}
\label{photoaff}
A&=\ln \bigg[\frac{k_{0L}k_{LR}k_{R0}}{k_{L0}k_{RL}k_{0R}}\bigg]\\ \notag &=-\beta[-e\Delta \Phi - \Delta E]-\beta_S\Delta E,
\end{align}
and the related entropy production rate $\dot{\sigma}$ determined by Eq.\ \eqref{eq3} yields
\begin{align}
\label{photoentropy}
\dot{\sigma}&=\langle J \rangle \bigg [-T^{-1}[-\Delta \Phi - \Delta E/e]-T_S^{-1}\Delta E/e \bigg].
\end{align}

In this cell model we assume that the "sun" is a monochromatic light source of energy $\Delta E$ such that the net heat flux from the "sun" is $\dot{\mathcal{Q}}_S=(\Delta E/e) \langle J \rangle$, where $\langle J \rangle$ is the average charge current in Eq.\ \eqref{photocurr}. The heat flux dissipated at the left and right electrode are $\dot{\mathcal{Q}}_L=(E_L/e - \Delta \Phi/2)\langle J \rangle$ and $\dot{\mathcal{Q}}_R=-(E_R/e + \Delta \Phi/2)\langle J \rangle$, respectively. Away from $V_{oc}$, at (nonequilibrium) steady state the power generated by the cell obtained from the first law as $P=-\Delta \Phi \langle J \rangle = \dot{\mathcal{Q}}_S+\dot{\mathcal{Q}}_L+\dot{\mathcal{Q}}_R$ ($P<0$ for $V_{oc}>\Delta \Phi>0$). At steady state, the entropy production rate is equal to the entropy flux into the thermal environments $\dot{\sigma}=-[\dot{\mathcal{Q}}_S/T_S+\dot{\mathcal{Q}}_L/T+\dot{\mathcal{Q}}_R/T]$.

\begin{figure}[h!!!!!]
\centering
\includegraphics[width=\linewidth]{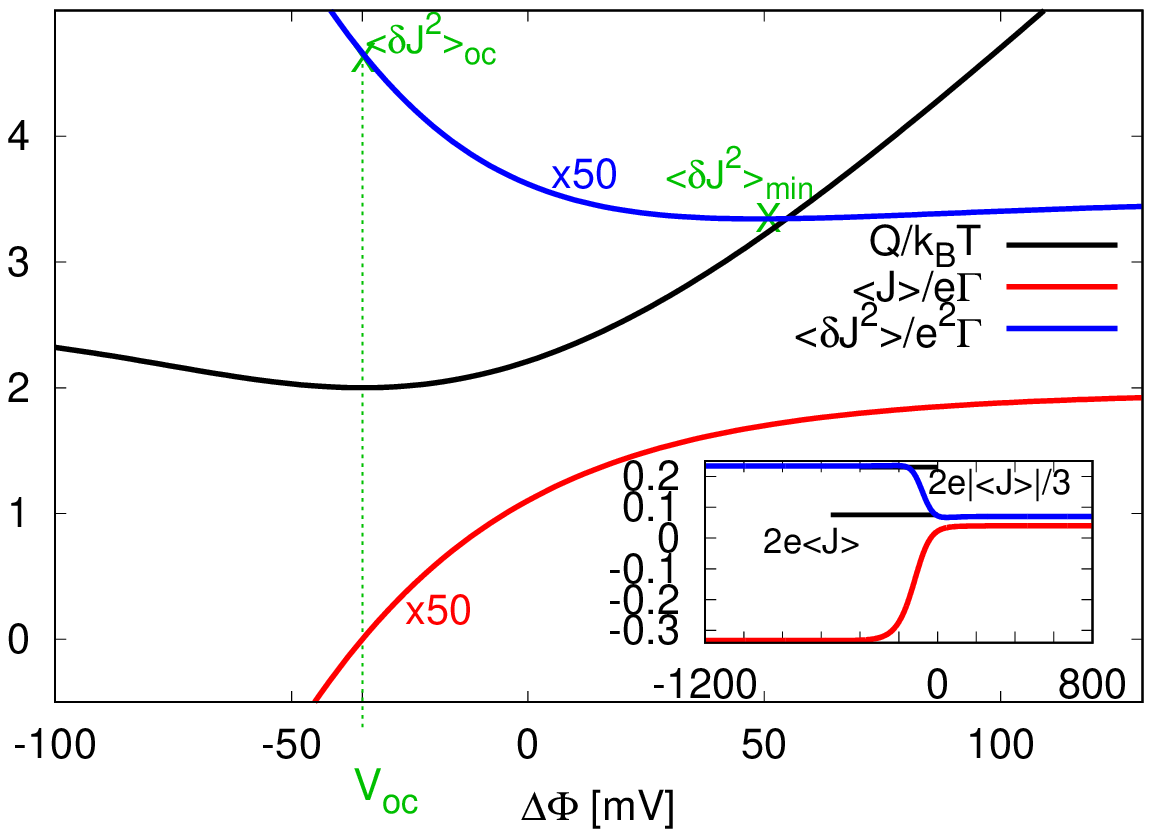}
\caption{\label{fig3c} $ Q/k_BT = \dot{\sigma} k_B^{-1}\langle \delta J^2 \rangle/ 2\langle J\rangle ^2$ (black), stationary charge current $\langle J \rangle$ (red) and the zero-frequency noise $\langle\delta J^2 \rangle$ (blue) against the applied bias potential $\Delta \Phi$. Inset: $\langle J \rangle$ and $\langle\delta J^2 \rangle$ against a larger range of applied bias potential $\Delta \Phi$. The temperature is $T=300$K and $k_BT_K= 26$meV, the "sun" temperature $T_S=461$K and $k_BT_S=40$meV. The energy difference of the upper and lower level is chosen to be $\Delta E=100$meV=$100\hbar \Gamma$ where $\Gamma$ is the electron transfer rate to each lead while we set $\hbar \equiv 1$. We choose the transfer rate between the two energy levels to $\Gamma_S=\Gamma$. The open circuit voltage is $V_{oc}=-(1-T/T_S)\Delta E/e=-35$mV.}
\end{figure}

Fig.\ \ref{fig3c} shows the characteristic behavior for photovoltaic devices. 
The scaled current $\langle J \rangle /e\Gamma$ is zero at the stopping voltage $ V_{oc}=-(1-T/T_S)\Delta E/e$ (here $V_{oc}=-35$meV), which is determined by the condition that the two drivings associated with the voltage $\Delta \Phi$ and with the thermal "sun" light (determined by the difference between $T$ and $T_S$) balance each other. For $\Delta \Phi > V_{oc}$ and $\Delta \Phi < V_{oc}$ the electronic  current flows in positive (left to right) and negative direction (right to left) respectively. The current becomes voltage independent for $\Delta \Phi - V_{oc}$ much smaller or larger than zero, where the saturation current on the negative $\Delta \Phi$ side exceeds that one on the positive side as another manifestation of the presence of two driving mechanisms that can join or oppose each other (see also later discussion on shot noise). 

Next consider the zero-frequency current noise at steady state in these voltage regimes. At the stopping voltage, $\Delta \Phi \to V_{oc}$, the current noise at zero-frequency has the same form as the Johnson-Nyquist result $\langle \delta J^2 \rangle_{oc}=4k_BT G_{oc}$, however $G_{oc}=d \langle J \rangle/ d\Delta \Phi|_{\Delta \Phi = V_{oc}}$ is a function of voltage and the two temperatures. Notably, in this truly nonequilibrium situation of two drivings $\langle \delta J^2 \rangle_{oc}$ is not the minimum noise as usually expected at zero current in absence of a driving (or, like in the present case, when both effectively compete each other) as in Sec.\ \ref{SecI}. We find $\langle \delta J^2 \rangle_{min}$ at $\Delta \Phi=48$mV for our choice of parameters (see Fig. \ref{fig3c}). 

Away from the stopping voltage, the shot noise is reached in two limiting cases when (a) $|-e\Delta\Phi-eV_{oc}| \gg k_BT$ or when (b) $e\Delta\Phi-eV_{oc} \gg k_BT$. Interestingly, the limiting shot noise values are different in the two limits. In case (a), where the hops $R\to E_R \to E_L \to L$ are all downhill, we find in the limit $\Delta E \gg k_BT_S$ (strongly reduced transfer from $E_L$ to $E_R$) $\langle \delta J^2 \rangle=2e|\langle J \rangle| [1-2(2\Gamma+\Gamma_S)\Gamma^2\Gamma_S/(2\Gamma_S\Gamma+\Gamma^2)^2]$.
The shot noise has the value $\langle \delta J^2 \rangle=2 e|\langle J \rangle|/3$ when $\Gamma=\Gamma_S$ (see inset in Fig.\ \ref{fig3c}) and is reduced in comparison the the classical noise obtained by Shottky by a factor of $3$. In case (b), when the hops $L \to E_L$ and $E_R \to R$ are downhill while the hop $E_L \to E_R$ is uphill, the total rate is dominated by the single $E_L \to E_R$ hop and we obtain the value of the classical shot noise \citep{sho1918} $\langle \delta J^2 \rangle=2 e\langle J \rangle$ (see inset in Fig.\ \ref{fig3c}) which is independent of the rates $\Gamma$ and $\Gamma_S$. 
The reduction of the classical shot noise for downhill transfer stems from the fact of two intermediate levels between the left and right electrode which is reminiscent of a "correlated" electron transfer as discussed in Sec.\ \ref{SecI} for one intermediate level with a maximal reduction factor of $2$.
For the uphill transfer, however, the transfer is strongly reduced and the shot noise shows its classical limit $2e\langle J \rangle$ like in charge transmission between two electrodes without intermediate state. 

Interestingly, when considering the TUR product normalized by temperature (defined by Eq.\ \eqref{TUR2}) $ Q/k_BT = \dot{\sigma} k_B^{-1}\langle \delta J^2 \rangle/2 \langle J\rangle^2$ (where $\dot{\sigma}$, $\langle \delta J^2 \rangle$ and $\langle J\rangle$ are taken from Eqs.\ \eqref{photoentropy}, \eqref{photonoise} and \eqref{photocurr} respectively), $Q/k_BT\geq 2$ while equality is reached when $\Delta\Phi=V_{oc}$ (see Appendix \ref{sec8}). Note that since the TUR is satisfied for the charge current a similar TUR inequality holds also for the energy current associated with photon absorption and emission where the current and noise have to be multiplied by the energy $\Delta E$ (or squared) which cancels each in the TUR product $Q$ while the entropy production remains the same (see also Ref.\ \cite{bar2015} for related currents in multicycle networks).

From a thermodynamic viewpoint, the solar cell is a heat engine converting heat input from a hot reservoir, the "sun", into work by moving electrons from a lower to a higher chemical potential against an applied bias potential. We now determine the thermodynamic efficiency at which energy conversion from sun photons to useful work takes place \cite{footnote2}. The thermodynamic efficiency reads \cite{esp2009,pie2016}
\begin{align}
\label{eff1}
\eta\equiv \frac{-\langle J \rangle \Delta \Phi}{\dot{\mathcal{Q}}_S}=\frac{-\langle J \rangle \Delta \Phi \eta_C}{T \dot{\sigma}-\langle J \rangle \Delta \Phi},
\end{align}
where we have used the relation $\dot{\mathcal{Q}}_S=(T \dot{\sigma}-\langle J \rangle\Delta \Phi)/\eta_C$ with the Carnot efficiency $\eta_C=1-T/T_S$.

Often the "internal" structure of the engine, i.e., the actual band gap $\Delta E$ and the resulting heat flows are difficult to determine. Following the consideration of Ref.\ \cite{pie2016} and given the uncertainty relation (Eq.\ \eqref{TUR2}), we can determine a bound for the entropy production rate, $\dot{\sigma}\geq 4 k_B \langle J \rangle ^2/ \langle \delta J^2 \rangle$, and find a bound for the heat supplied by the sun by $\dot{\mathcal{Q}}_S=(T \dot{\sigma}-\langle J \rangle \Delta \Phi)/\eta_C \geq (4 k_B T \langle J \rangle ^2/ \langle \delta J^2 \rangle-\langle J \rangle \Delta \Phi)/\eta_C$. 

This relation inserted in Eq.\ \eqref{eff1} yields an upper bound for the efficiency
\begin{align}
\label{eff2}
\eta \leq \eta_{TUR} &= \frac{-\langle J \rangle \Delta \Phi \eta_C}{4k_B T\langle J \rangle ^2/ \langle \delta J^2 \rangle-\langle J \rangle \Delta \Phi} \\ \notag
&= \frac{\eta_C}{4k_B Tg/ \langle \delta J^2 \rangle+1},
\end{align}
involving experimentally accessible quantities such as the average current $\langle J \rangle$, its zero-frequency noise $\langle \delta J^2 \rangle$ and the temperature of the leads $T$ only. For the last equality in Eq.\ \eqref{eff2} we use the definition $g=-\langle J \rangle/\Delta \Phi$, \cite{footnote3}.

\begin{figure}[h!!!!!]
\centering
\includegraphics[width=\linewidth]{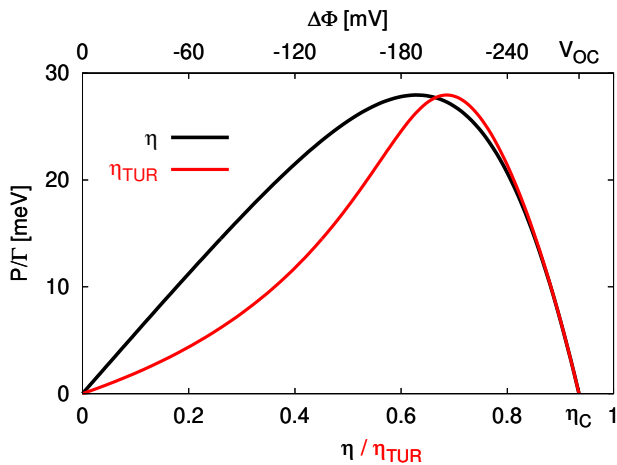}
\caption{\label{fig3e} Output power $P=-\langle J \rangle \Delta \Phi$ against thermodynamic efficiency $\eta$ (black) and the predicted bound on efficiency $\eta_{TUR}$ using TUR (red). The temperature is $T=300$K and $k_BT= 26$meV, the "sun" temperature $T_S=4610$K and $k_BT_S=400$meV. The Carnot efficiency for the photovoltaic cell $1-T/T_S=0.935$. The open circuit voltage is $V_{oc}=-280.5$mV. The energy difference of the upper and lower level is chosen to be $\Delta E=300$meV=$300\hbar \Gamma$ where $\Gamma$ is the electron transfer rate to each lead while we set $\hbar \equiv 1$. We choose the transfer rate between the two energy levels to $\Gamma_S=\Gamma$.}
\end{figure}

The power output of the photovoltaic cell $P=-\langle J\rangle \Delta \Phi$ ranges between zero, when either the average current or the applied bias potential vanishes, and a maximal value ($P>0$ for $V_{oc}<\Delta \Phi<0$). 
The efficiency at maximum power is larger under TUR equality (red line in Fig.\ \ref{fig3e}) than for the actual thermodynamic efficiency. 
The maximal efficiency performed by the cell is the Carnot efficiency $\eta_C=1-T/T_S$ (see Fig.\ \ref{fig3e}). In this case the stopping voltage $\Delta \Phi =V_{OC}$ is reached, the charge current vanishes and no power will be generated. This is reminiscent of a Carnot heat engine which operates adiabatically slowly and produces no power \citep{CarnotBook}. Going down in $|\Delta \Phi|$ below the stopping voltage (and hence decreasing $\eta$ from $\eta_C$) we see a linear response like regime where the increasing power remains only slightly below the TUR value. At the point of maximum power $\eta_{TUR}>\eta$. Away from this point, when operating at low $\Delta \Phi$ and therefore at low $\eta$, we see that one can achieve powers higher than the TUR value. 

By experimentally accessible quantities, i.e., temperatures of electrodes and "sun", the average current and its noise, one can estimate the efficiency of energy conversion without knowing the energy or band gap $\Delta E$ of the photovoltaic cell.

\section{Conclusion}
\label{conc}
We have investigated fluctuations, thermodynamic properties and possible bounds of realistic electrochemical junctions based on sequential hopping kinetics described by classical master equations. The zero-frequency noise in a metal-molecule-metal junction studied over the whole rage of potential bias shows two typical limits: The Johnson-Nynquist thermal noise for $e\Delta\Phi \ll k_BT$ and the shot noise for $e\Delta\Phi \gg k_BT$.
Interestingly, we could identify a component in the noise which can be associated to electron correlation, in analog to the quantum coherent tunneling, but arising here from the fact that a full charge transmission from one electron to the other always passes via an intermediate state. 
For equal transmission rates between molecule and both electrodes this correlation term maximizes and reduces the classical shot noise result by a factor of $2$. In general, we see that the classical correlation term arises from the assumption that only single site occupation is allowed. In "classical" junctions this may be due to strong Coulomb repulsion between charge carriers.
In the Marcus regime the solvent induced stabilization by the reorganization energy $E_R$ and the molecular level $\epsilon_d$ (reflecting the effect of a gate potential) strongly influences the thermal noise but have no impact on the shot noise. 
When considering the TUR in such an electrochemical junction its bound by $2k_BT$ is always satisfied in the equilibrium condition (zero potential bias). An increasing reorganization energy $E_r$ and molecular level $\epsilon_d$ brings the TUR even stronger above this bound when a potential bias is applied. Interestingly for $\epsilon_d=0$, uniform temperature of electrodes and solvent and symmetric coupling between molecule and both leads, we found an analytic expression for the TUR which serves minimal bound for a finite applied voltage. 
For a prototype photovoltaic cell the TUR bound of $2k_BT$ is reached when applying the stopping voltage where both competing drivings, incoming sun photons pushing charges "uphill a barrier" and the applied bias, compensate each other. However, the zero-frequency current noise goes through a minimum as function of the bias at an applied bias different from the stopping voltage, reflecting the true nonequilibrium due to the presence of two drivings. Interestingly, the shot noise is smaller by a factor of $3$ in comparison to Shottky noise when the two drivings by sun and applied voltage lump together.
Notably, the efficiency of energy conversion in the cell predicted by the relative current fluctuation together with the minimal TUR provide a satisfying upper bound for the thermodynamic efficiency. 
Extensions to nano heat engines governed by electron transmission processes unifying kinetic hopping and coherent tunneling (partial dephasing processes), see possible theoretical descriptions in Ref.\ \cite{sow2018,sow2019}, should be possible and are left for future research. We defer for a future study the consideration of further reducing noise if more complex system configurations are used.


\section*{Acknowledgments}
This work has been supported the U.S. National Science Foundation under the Grant No. CHE1953701 and the University of Pennsylvania. 

\section*{Data Availability}
The data that support the findings of this study are available from the corresponding author upon reasonable request.

\section*{References}
\bibliography{TUR}

\begin{thebibliography}{61}%
\makeatletter
\providecommand \@ifxundefined [1]{%
 \@ifx{#1\undefined}
}%
\providecommand \@ifnum [1]{%
 \ifnum #1\expandafter \@firstoftwo
 \else \expandafter \@secondoftwo
 \fi
}%
\providecommand \@ifx [1]{%
 \ifx #1\expandafter \@firstoftwo
 \else \expandafter \@secondoftwo
 \fi
}%
\providecommand \natexlab [1]{#1}%
\providecommand \enquote  [1]{``#1''}%
\providecommand \bibnamefont  [1]{#1}%
\providecommand \bibfnamefont [1]{#1}%
\providecommand \citenamefont [1]{#1}%
\providecommand \href@noop [0]{\@secondoftwo}%
\providecommand \href [0]{\begingroup \@sanitize@url \@href}%
\providecommand \@href[1]{\@@startlink{#1}\@@href}%
\providecommand \@@href[1]{\endgroup#1\@@endlink}%
\providecommand \@sanitize@url [0]{\catcode `\\12\catcode `\$12\catcode
  `\&12\catcode `\#12\catcode `\^12\catcode `\_12\catcode `\%12\relax}%
\providecommand \@@startlink[1]{}%
\providecommand \@@endlink[0]{}%
\providecommand \url  [0]{\begingroup\@sanitize@url \@url }%
\providecommand \@url [1]{\endgroup\@href {#1}{\urlprefix }}%
\providecommand \urlprefix  [0]{URL }%
\providecommand \Eprint [0]{\href }%
\providecommand \doibase [0]{http://dx.doi.org/}%
\providecommand \selectlanguage [0]{\@gobble}%
\providecommand \bibinfo  [0]{\@secondoftwo}%
\providecommand \bibfield  [0]{\@secondoftwo}%
\providecommand \translation [1]{[#1]}%
\providecommand \BibitemOpen [0]{}%
\providecommand \bibitemStop [0]{}%
\providecommand \bibitemNoStop [0]{.\EOS\space}%
\providecommand \EOS [0]{\spacefactor3000\relax}%
\providecommand \BibitemShut  [1]{\csname bibitem#1\endcsname}%
\let\auto@bib@innerbib\@empty
\bibitem [{\citenamefont {Carnot}(1824)}]{CarnotBook}%
  \BibitemOpen
  \bibfield  {author} {\bibinfo {author} {\bibfnamefont {S.}~\bibnamefont
  {Carnot}},\ }\href@noop {} {\emph {\bibinfo {title} {Réflexions sur la
  puissance motrice du feu et sur les machines propres à développer cette
  puissance}}}\ (\bibinfo  {publisher} {Bachelier},\ \bibinfo {address}
  {Paris},\ \bibinfo {year} {1824})\BibitemShut {NoStop}%
\bibitem [{\citenamefont {Clausius}(1854)}]{cla1854}%
  \BibitemOpen
  \bibfield  {author} {\bibinfo {author} {\bibfnamefont {R.}~\bibnamefont
  {Clausius}},\ }\href {\doibase https://doi.org/10.1002/andp.18541691202}
  {\bibfield  {journal} {\bibinfo  {journal} {Ann. Phys. Chem.}\ }\textbf
  {\bibinfo {volume} {93}},\ \bibinfo {pages} {81} (\bibinfo {year}
  {1854})}\BibitemShut {NoStop}%
\bibitem [{\citenamefont {Benenti}\ \emph {et~al.}(2017)\citenamefont
  {Benenti}, \citenamefont {Casati}, \citenamefont {Saito},\ and\ \citenamefont
  {Whitney}}]{ben2017}%
  \BibitemOpen
  \bibfield  {author} {\bibinfo {author} {\bibfnamefont {G.}~\bibnamefont
  {Benenti}}, \bibinfo {author} {\bibfnamefont {G.}~\bibnamefont {Casati}},
  \bibinfo {author} {\bibfnamefont {K.}~\bibnamefont {Saito}}, \ and\ \bibinfo
  {author} {\bibfnamefont {R.~S.}\ \bibnamefont {Whitney}},\ }\href {\doibase
  https://doi.org/10.1016/j.physrep.2017.05.008} {\bibfield  {journal}
  {\bibinfo  {journal} {Physics Reports}\ }\textbf {\bibinfo {volume} {694}},\
  \bibinfo {pages} {1} (\bibinfo {year} {2017})}\BibitemShut {NoStop}%
\bibitem [{\citenamefont {Jarzynski}(1997)}]{jar1997}%
  \BibitemOpen
  \bibfield  {author} {\bibinfo {author} {\bibfnamefont {C.}~\bibnamefont
  {Jarzynski}},\ }\href {\doibase 10.1103/PhysRevLett.78.2690} {\bibfield
  {journal} {\bibinfo  {journal} {Phys. Rev. Lett.}\ }\textbf {\bibinfo
  {volume} {78}},\ \bibinfo {pages} {2690} (\bibinfo {year}
  {1997})}\BibitemShut {NoStop}%
\bibitem [{\citenamefont {Seifert}(2012)}]{sei2012}%
  \BibitemOpen
  \bibfield  {author} {\bibinfo {author} {\bibfnamefont {U.}~\bibnamefont
  {Seifert}},\ }\href {\doibase 10.1088/0034-4885/75/12/126001} {\bibfield
  {journal} {\bibinfo  {journal} {Reports on Progress in Physics}\ }\textbf
  {\bibinfo {volume} {75}},\ \bibinfo {pages} {126001} (\bibinfo {year}
  {2012})}\BibitemShut {NoStop}%
\bibitem [{\citenamefont {{Van den Broeck}}\ and\ \citenamefont
  {Esposito}(2015)}]{van2015}%
  \BibitemOpen
  \bibfield  {author} {\bibinfo {author} {\bibfnamefont {C.}~\bibnamefont {{Van
  den Broeck}}}\ and\ \bibinfo {author} {\bibfnamefont {M.}~\bibnamefont
  {Esposito}},\ }\href {\doibase https://doi.org/10.1016/j.physa.2014.04.035}
  {\bibfield  {journal} {\bibinfo  {journal} {Physica A: Statistical Mechanics
  and its Applications}\ }\textbf {\bibinfo {volume} {418}},\ \bibinfo {pages}
  {6} (\bibinfo {year} {2015})}\BibitemShut {NoStop}%
\bibitem [{\citenamefont {Esposito}\ \emph {et~al.}(2009)\citenamefont
  {Esposito}, \citenamefont {Harbola},\ and\ \citenamefont
  {Mukamel}}]{esp2009}%
  \BibitemOpen
  \bibfield  {author} {\bibinfo {author} {\bibfnamefont {M.}~\bibnamefont
  {Esposito}}, \bibinfo {author} {\bibfnamefont {U.}~\bibnamefont {Harbola}}, \
  and\ \bibinfo {author} {\bibfnamefont {S.}~\bibnamefont {Mukamel}},\ }\href
  {\doibase 10.1103/RevModPhys.81.1665} {\bibfield  {journal} {\bibinfo
  {journal} {Rev. Mod. Phys.}\ }\textbf {\bibinfo {volume} {81}},\ \bibinfo
  {pages} {1665} (\bibinfo {year} {2009})}\BibitemShut {NoStop}%
\bibitem [{\citenamefont {Seifert}(2019)}]{sei2019}%
  \BibitemOpen
  \bibfield  {author} {\bibinfo {author} {\bibfnamefont {U.}~\bibnamefont
  {Seifert}},\ }\href {\doibase 10.1146/annurev-conmatphys-031218-013554}
  {\bibfield  {journal} {\bibinfo  {journal} {Annual Review of Condensed Matter
  Physics}\ }\textbf {\bibinfo {volume} {10}},\ \bibinfo {pages} {171}
  (\bibinfo {year} {2019})}\BibitemShut {NoStop}%
\bibitem [{\citenamefont {Barato}\ and\ \citenamefont
  {Seifert}(2015)}]{bar2015}%
  \BibitemOpen
  \bibfield  {author} {\bibinfo {author} {\bibfnamefont {A.~C.}\ \bibnamefont
  {Barato}}\ and\ \bibinfo {author} {\bibfnamefont {U.}~\bibnamefont
  {Seifert}},\ }\href {\doibase 10.1103/PhysRevLett.114.158101} {\bibfield
  {journal} {\bibinfo  {journal} {Phys. Rev. Lett.}\ }\textbf {\bibinfo
  {volume} {114}},\ \bibinfo {pages} {158101} (\bibinfo {year}
  {2015})}\BibitemShut {NoStop}%
\bibitem [{\citenamefont {Gingrich}\ \emph {et~al.}(2016)\citenamefont
  {Gingrich}, \citenamefont {Horowitz}, \citenamefont {Perunov},\ and\
  \citenamefont {England}}]{gin2016}%
  \BibitemOpen
  \bibfield  {author} {\bibinfo {author} {\bibfnamefont {T.~R.}\ \bibnamefont
  {Gingrich}}, \bibinfo {author} {\bibfnamefont {J.~M.}\ \bibnamefont
  {Horowitz}}, \bibinfo {author} {\bibfnamefont {N.}~\bibnamefont {Perunov}}, \
  and\ \bibinfo {author} {\bibfnamefont {J.~L.}\ \bibnamefont {England}},\
  }\href {\doibase 10.1103/PhysRevLett.116.120601} {\bibfield  {journal}
  {\bibinfo  {journal} {Phys. Rev. Lett.}\ }\textbf {\bibinfo {volume} {116}},\
  \bibinfo {pages} {120601} (\bibinfo {year} {2016})}\BibitemShut {NoStop}%
\bibitem [{\citenamefont {Horowitz}\ and\ \citenamefont
  {Gingrich}(2020)}]{hor2020}%
  \BibitemOpen
  \bibfield  {author} {\bibinfo {author} {\bibfnamefont {J.}~\bibnamefont
  {Horowitz}}\ and\ \bibinfo {author} {\bibfnamefont {T.}~\bibnamefont
  {Gingrich}},\ }\href {\doibase 10.1038/s41567-019-0702-6} {\bibfield
  {journal} {\bibinfo  {journal} {Nat. Phys.}\ ,\ \bibinfo {pages} {15}}
  (\bibinfo {year} {2020})}\BibitemShut {NoStop}%
\bibitem [{foo({\natexlab{a}})}]{footnote1}%
  \BibitemOpen
  \href@noop {} {\emph {\bibinfo {title} {Note that $\langle \delta J^2
  \rangle$ has the dimensionality charge squared per unit time. Usually the TUR
  is defined by $2\dot{\sigma}D / J^2$. These quanteties are related to our
  notation of average charge current by $\langle J \rangle=eJ$ and charge
  current noise at zero frequency by $\langle \delta J^2 \rangle=2e^2(\langle
  n^2 \rangle -\langle n \rangle^2)/t=4e^2D$}}}\BibitemShut {NoStop}%
\bibitem [{\citenamefont {Pietzonka}\ \emph {et~al.}(2016)\citenamefont
  {Pietzonka}, \citenamefont {Barato},\ and\ \citenamefont
  {Seifert}}]{pie2016}%
  \BibitemOpen
  \bibfield  {author} {\bibinfo {author} {\bibfnamefont {P.}~\bibnamefont
  {Pietzonka}}, \bibinfo {author} {\bibfnamefont {A.~C.}\ \bibnamefont
  {Barato}}, \ and\ \bibinfo {author} {\bibfnamefont {U.}~\bibnamefont
  {Seifert}},\ }\href {\doibase 10.1088/1742-5468/2016/12/124004} {\bibfield
  {journal} {\bibinfo  {journal} {Journal of Statistical Mechanics: Theory and
  Experiment}\ }\textbf {\bibinfo {volume} {2016}},\ \bibinfo {pages} {124004}
  (\bibinfo {year} {2016})}\BibitemShut {NoStop}%
\bibitem [{\citenamefont {Pietzonka}\ and\ \citenamefont
  {Seifert}(2018)}]{pie2018}%
  \BibitemOpen
  \bibfield  {author} {\bibinfo {author} {\bibfnamefont {P.}~\bibnamefont
  {Pietzonka}}\ and\ \bibinfo {author} {\bibfnamefont {U.}~\bibnamefont
  {Seifert}},\ }\href {\doibase 10.1103/PhysRevLett.120.190602} {\bibfield
  {journal} {\bibinfo  {journal} {Phys. Rev. Lett.}\ }\textbf {\bibinfo
  {volume} {120}},\ \bibinfo {pages} {190602} (\bibinfo {year}
  {2018})}\BibitemShut {NoStop}%
\bibitem [{\citenamefont {Jack}\ \emph {et~al.}(2020)\citenamefont {Jack},
  \citenamefont {L\'opez-Alamilla},\ and\ \citenamefont {Challis}}]{jac2020}%
  \BibitemOpen
  \bibfield  {author} {\bibinfo {author} {\bibfnamefont {M.~W.}\ \bibnamefont
  {Jack}}, \bibinfo {author} {\bibfnamefont {N.~J.}\ \bibnamefont
  {L\'opez-Alamilla}}, \ and\ \bibinfo {author} {\bibfnamefont {K.~J.}\
  \bibnamefont {Challis}},\ }\href {\doibase 10.1103/PhysRevE.101.062123}
  {\bibfield  {journal} {\bibinfo  {journal} {Phys. Rev. E}\ }\textbf {\bibinfo
  {volume} {101}},\ \bibinfo {pages} {062123} (\bibinfo {year}
  {2020})}\BibitemShut {NoStop}%
\bibitem [{\citenamefont {Saryal}\ \emph {et~al.}(2019)\citenamefont {Saryal},
  \citenamefont {Friedman}, \citenamefont {Segal},\ and\ \citenamefont
  {Agarwalla}}]{sar2019}%
  \BibitemOpen
  \bibfield  {author} {\bibinfo {author} {\bibfnamefont {S.}~\bibnamefont
  {Saryal}}, \bibinfo {author} {\bibfnamefont {H.~M.}\ \bibnamefont
  {Friedman}}, \bibinfo {author} {\bibfnamefont {D.}~\bibnamefont {Segal}}, \
  and\ \bibinfo {author} {\bibfnamefont {B.~K.}\ \bibnamefont {Agarwalla}},\
  }\href {\doibase 10.1103/PhysRevE.100.042101} {\bibfield  {journal} {\bibinfo
   {journal} {Phys. Rev. E}\ }\textbf {\bibinfo {volume} {100}},\ \bibinfo
  {pages} {042101} (\bibinfo {year} {2019})}\BibitemShut {NoStop}%
\bibitem [{\citenamefont {Barato}\ and\ \citenamefont
  {Seifert}(2016)}]{bar2016}%
  \BibitemOpen
  \bibfield  {author} {\bibinfo {author} {\bibfnamefont {A.~C.}\ \bibnamefont
  {Barato}}\ and\ \bibinfo {author} {\bibfnamefont {U.}~\bibnamefont
  {Seifert}},\ }\href {\doibase 10.1103/PhysRevX.6.041053} {\bibfield
  {journal} {\bibinfo  {journal} {Phys. Rev. X}\ }\textbf {\bibinfo {volume}
  {6}},\ \bibinfo {pages} {041053} (\bibinfo {year} {2016})}\BibitemShut
  {NoStop}%
\bibitem [{\citenamefont {Pal}\ \emph {et~al.}(2021)\citenamefont {Pal},
  \citenamefont {Reuveni},\ and\ \citenamefont {Rahav}}]{pal2021}%
  \BibitemOpen
  \bibfield  {author} {\bibinfo {author} {\bibfnamefont {A.}~\bibnamefont
  {Pal}}, \bibinfo {author} {\bibfnamefont {S.}~\bibnamefont {Reuveni}}, \ and\
  \bibinfo {author} {\bibfnamefont {S.}~\bibnamefont {Rahav}},\ }\href
  {\doibase 10.1103/PhysRevResearch.3.013273} {\bibfield  {journal} {\bibinfo
  {journal} {Phys. Rev. Research}\ }\textbf {\bibinfo {volume} {3}},\ \bibinfo
  {pages} {013273} (\bibinfo {year} {2021})}\BibitemShut {NoStop}%
\bibitem [{\citenamefont {Pietzonka}(2022)}]{pie2022}%
  \BibitemOpen
  \bibfield  {author} {\bibinfo {author} {\bibfnamefont {P.}~\bibnamefont
  {Pietzonka}},\ }\href {\doibase 10.1103/PhysRevLett.128.130606} {\bibfield
  {journal} {\bibinfo  {journal} {Phys. Rev. Lett.}\ }\textbf {\bibinfo
  {volume} {128}},\ \bibinfo {pages} {130606} (\bibinfo {year}
  {2022})}\BibitemShut {NoStop}%
\bibitem [{\citenamefont {Cangemi}\ \emph {et~al.}(2020)\citenamefont
  {Cangemi}, \citenamefont {Cataudella}, \citenamefont {Benenti}, \citenamefont
  {Sassetti},\ and\ \citenamefont {De~Filippis}}]{can2020}%
  \BibitemOpen
  \bibfield  {author} {\bibinfo {author} {\bibfnamefont {L.~M.}\ \bibnamefont
  {Cangemi}}, \bibinfo {author} {\bibfnamefont {V.}~\bibnamefont {Cataudella}},
  \bibinfo {author} {\bibfnamefont {G.}~\bibnamefont {Benenti}}, \bibinfo
  {author} {\bibfnamefont {M.}~\bibnamefont {Sassetti}}, \ and\ \bibinfo
  {author} {\bibfnamefont {G.}~\bibnamefont {De~Filippis}},\ }\href {\doibase
  10.1103/PhysRevB.102.165418} {\bibfield  {journal} {\bibinfo  {journal}
  {Phys. Rev. B}\ }\textbf {\bibinfo {volume} {102}},\ \bibinfo {pages}
  {165418} (\bibinfo {year} {2020})}\BibitemShut {NoStop}%
\bibitem [{\citenamefont {Kalaee}\ \emph {et~al.}(2021)\citenamefont {Kalaee},
  \citenamefont {Wacker},\ and\ \citenamefont {Potts}}]{kal2021}%
  \BibitemOpen
  \bibfield  {author} {\bibinfo {author} {\bibfnamefont {A.~A.~S.}\
  \bibnamefont {Kalaee}}, \bibinfo {author} {\bibfnamefont {A.}~\bibnamefont
  {Wacker}}, \ and\ \bibinfo {author} {\bibfnamefont {P.~P.}\ \bibnamefont
  {Potts}},\ }\href {\doibase 10.1103/PhysRevE.104.L012103} {\bibfield
  {journal} {\bibinfo  {journal} {Phys. Rev. E}\ }\textbf {\bibinfo {volume}
  {104}},\ \bibinfo {pages} {L012103} (\bibinfo {year} {2021})}\BibitemShut
  {NoStop}%
\bibitem [{\citenamefont {Brandner}\ \emph {et~al.}(2018)\citenamefont
  {Brandner}, \citenamefont {Hanazato},\ and\ \citenamefont {Saito}}]{bra2018}%
  \BibitemOpen
  \bibfield  {author} {\bibinfo {author} {\bibfnamefont {K.}~\bibnamefont
  {Brandner}}, \bibinfo {author} {\bibfnamefont {T.}~\bibnamefont {Hanazato}},
  \ and\ \bibinfo {author} {\bibfnamefont {K.}~\bibnamefont {Saito}},\ }\href
  {\doibase 10.1103/PhysRevLett.120.090601} {\bibfield  {journal} {\bibinfo
  {journal} {Phys. Rev. Lett.}\ }\textbf {\bibinfo {volume} {120}},\ \bibinfo
  {pages} {090601} (\bibinfo {year} {2018})}\BibitemShut {NoStop}%
\bibitem [{\citenamefont {Agarwalla}\ and\ \citenamefont
  {Segal}(2018)}]{aga2018}%
  \BibitemOpen
  \bibfield  {author} {\bibinfo {author} {\bibfnamefont {B.~K.}\ \bibnamefont
  {Agarwalla}}\ and\ \bibinfo {author} {\bibfnamefont {D.}~\bibnamefont
  {Segal}},\ }\href {\doibase 10.1103/PhysRevB.98.155438} {\bibfield  {journal}
  {\bibinfo  {journal} {Phys. Rev. B}\ }\textbf {\bibinfo {volume} {98}},\
  \bibinfo {pages} {155438} (\bibinfo {year} {2018})}\BibitemShut {NoStop}%
\bibitem [{\citenamefont {Ptaszy\ifmmode~\acute{n}\else
  \'{n}\fi{}ski}(2018)}]{pta2018}%
  \BibitemOpen
  \bibfield  {author} {\bibinfo {author} {\bibfnamefont {K.}~\bibnamefont
  {Ptaszy\ifmmode~\acute{n}\else \'{n}\fi{}ski}},\ }\href {\doibase
  10.1103/PhysRevB.98.085425} {\bibfield  {journal} {\bibinfo  {journal} {Phys.
  Rev. B}\ }\textbf {\bibinfo {volume} {98}},\ \bibinfo {pages} {085425}
  (\bibinfo {year} {2018})}\BibitemShut {NoStop}%
\bibitem [{\citenamefont {Liu}\ and\ \citenamefont {Segal}(2019)}]{jun2019}%
  \BibitemOpen
  \bibfield  {author} {\bibinfo {author} {\bibfnamefont {J.}~\bibnamefont
  {Liu}}\ and\ \bibinfo {author} {\bibfnamefont {D.}~\bibnamefont {Segal}},\
  }\href {\doibase 10.1103/PhysRevE.99.062141} {\bibfield  {journal} {\bibinfo
  {journal} {Phys. Rev. E}\ }\textbf {\bibinfo {volume} {99}},\ \bibinfo
  {pages} {062141} (\bibinfo {year} {2019})}\BibitemShut {NoStop}%
\bibitem [{\citenamefont {Friedman}\ \emph {et~al.}(2020)\citenamefont
  {Friedman}, \citenamefont {Agarwalla}, \citenamefont {Shein-Lumbroso},
  \citenamefont {Tal},\ and\ \citenamefont {Segal}}]{fri2020}%
  \BibitemOpen
  \bibfield  {author} {\bibinfo {author} {\bibfnamefont {H.~M.}\ \bibnamefont
  {Friedman}}, \bibinfo {author} {\bibfnamefont {B.~K.}\ \bibnamefont
  {Agarwalla}}, \bibinfo {author} {\bibfnamefont {O.}~\bibnamefont
  {Shein-Lumbroso}}, \bibinfo {author} {\bibfnamefont {O.}~\bibnamefont {Tal}},
  \ and\ \bibinfo {author} {\bibfnamefont {D.}~\bibnamefont {Segal}},\ }\href
  {\doibase 10.1103/PhysRevB.101.195423} {\bibfield  {journal} {\bibinfo
  {journal} {Phys. Rev. B}\ }\textbf {\bibinfo {volume} {101}},\ \bibinfo
  {pages} {195423} (\bibinfo {year} {2020})}\BibitemShut {NoStop}%
\bibitem [{\citenamefont {Marcus}(1956{\natexlab{a}})}]{mar1956a}%
  \BibitemOpen
  \bibfield  {author} {\bibinfo {author} {\bibfnamefont {R.~A.}\ \bibnamefont
  {Marcus}},\ }\href {\doibase 10.1063/1.1742723} {\bibfield  {journal}
  {\bibinfo  {journal} {The Journal of Chemical Physics}\ }\textbf {\bibinfo
  {volume} {24}},\ \bibinfo {pages} {966} (\bibinfo {year}
  {1956}{\natexlab{a}})}\BibitemShut {NoStop}%
\bibitem [{\citenamefont {Marcus}(1956{\natexlab{b}})}]{mar1956b}%
  \BibitemOpen
  \bibfield  {author} {\bibinfo {author} {\bibfnamefont {R.~A.}\ \bibnamefont
  {Marcus}},\ }\href {\doibase 10.1063/1.1742724} {\bibfield  {journal}
  {\bibinfo  {journal} {The Journal of Chemical Physics}\ }\textbf {\bibinfo
  {volume} {24}},\ \bibinfo {pages} {979} (\bibinfo {year}
  {1956}{\natexlab{b}})}\BibitemShut {NoStop}%
\bibitem [{\citenamefont {Marcus}\ and\ \citenamefont {Sumi}(1986)}]{mar1986}%
  \BibitemOpen
  \bibfield  {author} {\bibinfo {author} {\bibfnamefont {R.}~\bibnamefont
  {Marcus}}\ and\ \bibinfo {author} {\bibfnamefont {H.}~\bibnamefont {Sumi}},\
  }\href {\doibase https://doi.org/10.1016/0022-0728(86)80507-1} {\bibfield
  {journal} {\bibinfo  {journal} {J. Electroanal. Chem.}\ }\textbf {\bibinfo
  {volume} {204}},\ \bibinfo {pages} {59} (\bibinfo {year} {1986})}\BibitemShut
  {NoStop}%
\bibitem [{\citenamefont {Zusman}(1980)}]{zus1980}%
  \BibitemOpen
  \bibfield  {author} {\bibinfo {author} {\bibfnamefont {L.}~\bibnamefont
  {Zusman}},\ }\href {\doibase https://doi.org/10.1016/0301-0104(80)85267-0}
  {\bibfield  {journal} {\bibinfo  {journal} {Chemical Physics}\ }\textbf
  {\bibinfo {volume} {49}},\ \bibinfo {pages} {295} (\bibinfo {year}
  {1980})}\BibitemShut {NoStop}%
\bibitem [{\citenamefont {Zusman}(1995)}]{zus1995}%
  \BibitemOpen
  \bibfield  {author} {\bibinfo {author} {\bibfnamefont {L.}~\bibnamefont
  {Zusman}},\ }\href {\doibase 10.1063/1.468688} {\bibfield  {journal}
  {\bibinfo  {journal} {The Journal of Chemical Physics}\ }\textbf {\bibinfo
  {volume} {102}},\ \bibinfo {pages} {2580} (\bibinfo {year}
  {1995})}\BibitemShut {NoStop}%
\bibitem [{\citenamefont {Zhang}\ \emph {et~al.}(2008)\citenamefont {Zhang},
  \citenamefont {Kuznetsov}, \citenamefont {Medvedev}, \citenamefont {Chi},
  \citenamefont {Albrecht}, \citenamefont {Jensen},\ and\ \citenamefont
  {Ulstrup}}]{zha2008}%
  \BibitemOpen
  \bibfield  {author} {\bibinfo {author} {\bibfnamefont {J.}~\bibnamefont
  {Zhang}}, \bibinfo {author} {\bibfnamefont {A.~M.}\ \bibnamefont
  {Kuznetsov}}, \bibinfo {author} {\bibfnamefont {I.~G.}\ \bibnamefont
  {Medvedev}}, \bibinfo {author} {\bibfnamefont {Q.}~\bibnamefont {Chi}},
  \bibinfo {author} {\bibfnamefont {T.}~\bibnamefont {Albrecht}}, \bibinfo
  {author} {\bibfnamefont {P.~S.}\ \bibnamefont {Jensen}}, \ and\ \bibinfo
  {author} {\bibfnamefont {J.}~\bibnamefont {Ulstrup}},\ }\href {\doibase
  10.1021/cr068073+} {\bibfield  {journal} {\bibinfo  {journal} {Chemical
  Reviews}\ }\textbf {\bibinfo {volume} {108}},\ \bibinfo {pages} {2737}
  (\bibinfo {year} {2008})}\BibitemShut {NoStop}%
\bibitem [{\citenamefont {Landau}(1932)}]{lan1932}%
  \BibitemOpen
  \bibfield  {author} {\bibinfo {author} {\bibfnamefont {L.}~\bibnamefont
  {Landau}},\ }\href@noop {} {\bibfield  {journal} {\bibinfo  {journal} {Phys Z
  Sowjetunion}\ }\textbf {\bibinfo {volume} {2}},\ \bibinfo {pages} {46}
  (\bibinfo {year} {1932})}\BibitemShut {NoStop}%
\bibitem [{\citenamefont {Blanter}\ and\ \citenamefont
  {Büttiker}(2000)}]{bla2000}%
  \BibitemOpen
  \bibfield  {author} {\bibinfo {author} {\bibfnamefont {Y.}~\bibnamefont
  {Blanter}}\ and\ \bibinfo {author} {\bibfnamefont {M.}~\bibnamefont
  {Büttiker}},\ }\href {\doibase
  https://doi.org/10.1016/S0370-1573(99)00123-4} {\bibfield  {journal}
  {\bibinfo  {journal} {Physics Reports}\ }\textbf {\bibinfo {volume} {336}},\
  \bibinfo {pages} {1} (\bibinfo {year} {2000})}\BibitemShut {NoStop}%
\bibitem [{\citenamefont {Galperin}\ and\ \citenamefont
  {Nitzan}(2005)}]{gal2005}%
  \BibitemOpen
  \bibfield  {author} {\bibinfo {author} {\bibfnamefont {M.}~\bibnamefont
  {Galperin}}\ and\ \bibinfo {author} {\bibfnamefont {A.}~\bibnamefont
  {Nitzan}},\ }\href {\doibase 10.1103/PhysRevLett.95.206802} {\bibfield
  {journal} {\bibinfo  {journal} {Phys. Rev. Lett.}\ }\textbf {\bibinfo
  {volume} {95}},\ \bibinfo {pages} {206802} (\bibinfo {year}
  {2005})}\BibitemShut {NoStop}%
\bibitem [{\citenamefont {Galperin}\ and\ \citenamefont
  {Nitzan}(2006)}]{gal2006}%
  \BibitemOpen
  \bibfield  {author} {\bibinfo {author} {\bibfnamefont {M.}~\bibnamefont
  {Galperin}}\ and\ \bibinfo {author} {\bibfnamefont {A.}~\bibnamefont
  {Nitzan}},\ }\href {\doibase 10.1063/1.2204917} {\bibfield  {journal}
  {\bibinfo  {journal} {The Journal of Chemical Physics}\ }\textbf {\bibinfo
  {volume} {124}},\ \bibinfo {pages} {234709} (\bibinfo {year}
  {2006})}\BibitemShut {NoStop}%
\bibitem [{\citenamefont {Jarzynski}(2011)}]{jar2011}%
  \BibitemOpen
  \bibfield  {author} {\bibinfo {author} {\bibfnamefont {C.}~\bibnamefont
  {Jarzynski}},\ }\href {\doibase 10.1146/annurev-conmatphys-062910-140506}
  {\bibfield  {journal} {\bibinfo  {journal} {Annual Review of Condensed Matter
  Physics}\ }\textbf {\bibinfo {volume} {2}},\ \bibinfo {pages} {329} (\bibinfo
  {year} {2011})}\BibitemShut {NoStop}%
\bibitem [{\citenamefont {Prigogine}(1955)}]{PrigogineBook}%
  \BibitemOpen
  \bibfield  {author} {\bibinfo {author} {\bibfnamefont {I.}~\bibnamefont
  {Prigogine}},\ }\href@noop {} {\emph {\bibinfo {title} {Introduction to
  Thermodynamics of Irreversible Processes}}}\ (\bibinfo  {publisher} {Charles
  C Thomas Publisher},\ \bibinfo {address} {Springfield},\ \bibinfo {year}
  {1955})\BibitemShut {NoStop}%
\bibitem [{\citenamefont {Tom\'e}\ and\ \citenamefont
  {de~Oliveira}(2012)}]{tom2005}%
  \BibitemOpen
  \bibfield  {author} {\bibinfo {author} {\bibfnamefont {T.}~\bibnamefont
  {Tom\'e}}\ and\ \bibinfo {author} {\bibfnamefont {M.~J.}\ \bibnamefont
  {de~Oliveira}},\ }\href {\doibase 10.1103/PhysRevLett.108.020601} {\bibfield
  {journal} {\bibinfo  {journal} {Phys. Rev. Lett.}\ }\textbf {\bibinfo
  {volume} {108}},\ \bibinfo {pages} {020601} (\bibinfo {year}
  {2012})}\BibitemShut {NoStop}%
\bibitem [{\citenamefont {Esposito}\ and\ \citenamefont {Van~den
  Broeck}(2010)}]{esp2010}%
  \BibitemOpen
  \bibfield  {author} {\bibinfo {author} {\bibfnamefont {M.}~\bibnamefont
  {Esposito}}\ and\ \bibinfo {author} {\bibfnamefont {C.}~\bibnamefont {Van~den
  Broeck}},\ }\href {\doibase 10.1103/PhysRevE.82.011143} {\bibfield  {journal}
  {\bibinfo  {journal} {Phys. Rev. E}\ }\textbf {\bibinfo {volume} {82}},\
  \bibinfo {pages} {011143} (\bibinfo {year} {2010})}\BibitemShut {NoStop}%
\bibitem [{\citenamefont {Koza}(1999)}]{koz1999}%
  \BibitemOpen
  \bibfield  {author} {\bibinfo {author} {\bibfnamefont {Z.}~\bibnamefont
  {Koza}},\ }\href {\doibase 10.1088/0305-4470/32/44/303} {\bibfield  {journal}
  {\bibinfo  {journal} {Journal of Physics A: Mathematical and General}\
  }\textbf {\bibinfo {volume} {32}},\ \bibinfo {pages} {7637} (\bibinfo {year}
  {1999})}\BibitemShut {NoStop}%
\bibitem [{\citenamefont {Koza}(2002)}]{koz2002}%
  \BibitemOpen
  \bibfield  {author} {\bibinfo {author} {\bibfnamefont {Z.}~\bibnamefont
  {Koza}},\ }\href {\doibase 10.1103/PhysRevE.65.031905} {\bibfield  {journal}
  {\bibinfo  {journal} {Phys. Rev. E}\ }\textbf {\bibinfo {volume} {65}},\
  \bibinfo {pages} {031905} (\bibinfo {year} {2002})}\BibitemShut {NoStop}%
\bibitem [{\citenamefont {B\"uttiker}(1990)}]{but1990}%
  \BibitemOpen
  \bibfield  {author} {\bibinfo {author} {\bibfnamefont {M.}~\bibnamefont
  {B\"uttiker}},\ }\href {\doibase 10.1103/PhysRevLett.65.2901} {\bibfield
  {journal} {\bibinfo  {journal} {Phys. Rev. Lett.}\ }\textbf {\bibinfo
  {volume} {65}},\ \bibinfo {pages} {2901} (\bibinfo {year}
  {1990})}\BibitemShut {NoStop}%
\bibitem [{\citenamefont {Schottky}(1918)}]{sho1918}%
  \BibitemOpen
  \bibfield  {author} {\bibinfo {author} {\bibfnamefont {W.}~\bibnamefont
  {Schottky}},\ }\href {\doibase https://doi.org/10.1002/andp.19183622304}
  {\bibfield  {journal} {\bibinfo  {journal} {Ann. Phys.}\ }\textbf {\bibinfo
  {volume} {57}},\ \bibinfo {pages} {541} (\bibinfo {year} {1918})}\BibitemShut
  {NoStop}%
\bibitem [{\citenamefont {Lesovik}(1989)}]{les1989}%
  \BibitemOpen
  \bibfield  {author} {\bibinfo {author} {\bibfnamefont {G.~B.}\ \bibnamefont
  {Lesovik}},\ }\href@noop {} {\bibfield  {journal} {\bibinfo  {journal} {JETP
  Lett.}\ }\textbf {\bibinfo {volume} {49}},\ \bibinfo {pages} {594} (\bibinfo
  {year} {1989})}\BibitemShut {NoStop}%
\bibitem [{\citenamefont {Johnson}(1927)}]{joh1927}%
  \BibitemOpen
  \bibfield  {author} {\bibinfo {author} {\bibfnamefont {J.}~\bibnamefont
  {Johnson}},\ }\href {\doibase https://doi.org/10.1038/119050c0} {\bibfield
  {journal} {\bibinfo  {journal} {Nature}\ }\textbf {\bibinfo {volume} {119}},\
  \bibinfo {pages} {50} (\bibinfo {year} {1927})}\BibitemShut {NoStop}%
\bibitem [{\citenamefont {Johnson}(1928)}]{joh1928}%
  \BibitemOpen
  \bibfield  {author} {\bibinfo {author} {\bibfnamefont {J.~B.}\ \bibnamefont
  {Johnson}},\ }\href {\doibase 10.1103/PhysRev.32.97} {\bibfield  {journal}
  {\bibinfo  {journal} {Phys. Rev.}\ }\textbf {\bibinfo {volume} {32}},\
  \bibinfo {pages} {97} (\bibinfo {year} {1928})}\BibitemShut {NoStop}%
\bibitem [{\citenamefont {Nyquist}(1928)}]{nyq1928}%
  \BibitemOpen
  \bibfield  {author} {\bibinfo {author} {\bibfnamefont {H.}~\bibnamefont
  {Nyquist}},\ }\href {\doibase 10.1103/PhysRev.32.110} {\bibfield  {journal}
  {\bibinfo  {journal} {Phys. Rev.}\ }\textbf {\bibinfo {volume} {32}},\
  \bibinfo {pages} {110} (\bibinfo {year} {1928})}\BibitemShut {NoStop}%
\bibitem [{\citenamefont {Migliore}\ \emph {et~al.}(2012)\citenamefont
  {Migliore}, \citenamefont {Schiff},\ and\ \citenamefont {Nitzan}}]{mig2012}%
  \BibitemOpen
  \bibfield  {author} {\bibinfo {author} {\bibfnamefont {A.}~\bibnamefont
  {Migliore}}, \bibinfo {author} {\bibfnamefont {P.}~\bibnamefont {Schiff}}, \
  and\ \bibinfo {author} {\bibfnamefont {A.}~\bibnamefont {Nitzan}},\ }\href
  {\doibase 10.1039/C2CP41442B} {\bibfield  {journal} {\bibinfo  {journal}
  {Phys. Chem. Chem. Phys.}\ }\textbf {\bibinfo {volume} {14}},\ \bibinfo
  {pages} {13746} (\bibinfo {year} {2012})}\BibitemShut {NoStop}%
\bibitem [{\citenamefont {Spietz}\ \emph {et~al.}(2003)\citenamefont {Spietz},
  \citenamefont {Lehnert}, \citenamefont {Siddiqi},\ and\ \citenamefont
  {Schoelkopf}}]{spi2003}%
  \BibitemOpen
  \bibfield  {author} {\bibinfo {author} {\bibfnamefont {L.}~\bibnamefont
  {Spietz}}, \bibinfo {author} {\bibfnamefont {K.~W.}\ \bibnamefont {Lehnert}},
  \bibinfo {author} {\bibfnamefont {I.}~\bibnamefont {Siddiqi}}, \ and\
  \bibinfo {author} {\bibfnamefont {R.~J.}\ \bibnamefont {Schoelkopf}},\ }\href
  {\doibase 10.1126/science.1084647} {\bibfield  {journal} {\bibinfo  {journal}
  {Science}\ }\textbf {\bibinfo {volume} {300}},\ \bibinfo {pages} {1929}
  (\bibinfo {year} {2003})}\BibitemShut {NoStop}%
\bibitem [{\citenamefont {Boltzmann}(1877)}]{bol1877}%
  \BibitemOpen
  \bibfield  {author} {\bibinfo {author} {\bibfnamefont {L.}~\bibnamefont
  {Boltzmann}},\ }\href@noop {} {\bibfield  {journal} {\bibinfo  {journal}
  {Sitz.-Ber. Akad. Wiss.}\ }\textbf {\bibinfo {volume} {75}},\ \bibinfo
  {pages} {67} (\bibinfo {year} {1877})}\BibitemShut {NoStop}%
\bibitem [{\citenamefont {Gibbs}(1981)}]{GibbsBook}%
  \BibitemOpen
  \bibfield  {author} {\bibinfo {author} {\bibfnamefont {J.~W.}\ \bibnamefont
  {Gibbs}},\ }\href@noop {} {\emph {\bibinfo {title} {Elementary Principles in
  Statistical Mechanics}}}\ (\bibinfo  {publisher} {Ox Bow Press},\ \bibinfo
  {address} {Woodbridge},\ \bibinfo {year} {1981})\BibitemShut {NoStop}%
\bibitem [{\citenamefont {Schnakenberg}(1976)}]{sna1976}%
  \BibitemOpen
  \bibfield  {author} {\bibinfo {author} {\bibfnamefont {J.}~\bibnamefont
  {Schnakenberg}},\ }\href {\doibase 10.1103/RevModPhys.48.571} {\bibfield
  {journal} {\bibinfo  {journal} {Rev. Mod. Phys.}\ }\textbf {\bibinfo {volume}
  {48}},\ \bibinfo {pages} {571} (\bibinfo {year} {1976})}\BibitemShut
  {NoStop}%
\bibitem [{\citenamefont {Dechant}\ and\ \citenamefont {Sasa}(2018)}]{dec2018}%
  \BibitemOpen
  \bibfield  {author} {\bibinfo {author} {\bibfnamefont {A.}~\bibnamefont
  {Dechant}}\ and\ \bibinfo {author} {\bibfnamefont {S.}~\bibnamefont {Sasa}},\
  }\href {\doibase 10.1088/1742-5468/aac91a} {\bibfield  {journal} {\bibinfo
  {journal} {Journal of Statistical Mechanics: Theory and Experiment}\ }\textbf
  {\bibinfo {volume} {2018}},\ \bibinfo {pages} {063209} (\bibinfo {year}
  {2018})}\BibitemShut {NoStop}%
\bibitem [{\citenamefont {Kirchberg}\ \emph {et~al.}(2020)\citenamefont
  {Kirchberg}, \citenamefont {Thorwart},\ and\ \citenamefont
  {Nitzan}}]{kir2020}%
  \BibitemOpen
  \bibfield  {author} {\bibinfo {author} {\bibfnamefont {H.}~\bibnamefont
  {Kirchberg}}, \bibinfo {author} {\bibfnamefont {M.}~\bibnamefont {Thorwart}},
  \ and\ \bibinfo {author} {\bibfnamefont {A.}~\bibnamefont {Nitzan}},\ }\href
  {\doibase 10.1021/acs.jpclett.0c00118} {\bibfield  {journal} {\bibinfo
  {journal} {The Journal of Physical Chemistry Letters}\ }\textbf {\bibinfo
  {volume} {11}},\ \bibinfo {pages} {1729} (\bibinfo {year}
  {2020})}\BibitemShut {NoStop}%
\bibitem [{\citenamefont {Kirchberg}\ and\ \citenamefont
  {Nitzan}(2022)}]{kir2022}%
  \BibitemOpen
  \bibfield  {author} {\bibinfo {author} {\bibfnamefont {H.}~\bibnamefont
  {Kirchberg}}\ and\ \bibinfo {author} {\bibfnamefont {A.}~\bibnamefont
  {Nitzan}},\ }\href {\doibase 10.1063/5.0086319} {\bibfield  {journal}
  {\bibinfo  {journal} {The Journal of Chemical Physics}\ }\textbf {\bibinfo
  {volume} {156}},\ \bibinfo {pages} {094306} (\bibinfo {year}
  {2022})}\BibitemShut {NoStop}%
\bibitem [{\citenamefont {Rutten}\ \emph {et~al.}(2009)\citenamefont {Rutten},
  \citenamefont {Esposito},\ and\ \citenamefont {Cleuren}}]{rut2009}%
  \BibitemOpen
  \bibfield  {author} {\bibinfo {author} {\bibfnamefont {B.}~\bibnamefont
  {Rutten}}, \bibinfo {author} {\bibfnamefont {M.}~\bibnamefont {Esposito}}, \
  and\ \bibinfo {author} {\bibfnamefont {B.}~\bibnamefont {Cleuren}},\ }\href
  {\doibase 10.1103/PhysRevB.80.235122} {\bibfield  {journal} {\bibinfo
  {journal} {Phys. Rev. B}\ }\textbf {\bibinfo {volume} {80}},\ \bibinfo
  {pages} {235122} (\bibinfo {year} {2009})}\BibitemShut {NoStop}%
\bibitem [{foo({\natexlab{b}})}]{footnote2}%
  \BibitemOpen
  \href@noop {} {\emph {\bibinfo {title} {As stated $\langle J \rangle$ is the
  electronic current, $\Delta \Phi$ negative means $e\Delta \Phi_R>0$ and
  postive work increases the energy of the device}}}\BibitemShut {NoStop}%
\bibitem [{foo({\natexlab{c}})}]{footnote3}%
  \BibitemOpen
  \href@noop {} {\emph {\bibinfo {title} {By evaluating the thermodynamic
  efficiency $\eta=-\langle J \rangle \Delta \Phi/\Delta E/e \langle J \rangle
  = \Delta \Phi/V_{oc}$ and using the inequality in Eq. (44), we can further
  formulate a lower bound for the stopping voltage $V_{oc}\geq 4k_BT \langle J
  \rangle/\langle \delta J^2 \rangle + \Delta \Phi$ and thus for $\Delta
  E$.}}}\BibitemShut {Stop}%
\bibitem [{\citenamefont {Sowa}\ \emph {et~al.}(2018)\citenamefont {Sowa},
  \citenamefont {Mol}, \citenamefont {Briggs},\ and\ \citenamefont
  {Gauger}}]{sow2018}%
  \BibitemOpen
  \bibfield  {author} {\bibinfo {author} {\bibfnamefont {J.~K.}\ \bibnamefont
  {Sowa}}, \bibinfo {author} {\bibfnamefont {J.~A.}\ \bibnamefont {Mol}},
  \bibinfo {author} {\bibfnamefont {G.~A.~D.}\ \bibnamefont {Briggs}}, \ and\
  \bibinfo {author} {\bibfnamefont {E.~M.}\ \bibnamefont {Gauger}},\ }\href
  {\doibase 10.1063/1.5049537} {\bibfield  {journal} {\bibinfo  {journal} {The
  Journal of Chemical Physics}\ }\textbf {\bibinfo {volume} {149}},\ \bibinfo
  {pages} {154112} (\bibinfo {year} {2018})}\BibitemShut {NoStop}%
\bibitem [{\citenamefont {Sowa}\ \emph {et~al.}(2019)\citenamefont {Sowa},
  \citenamefont {Mol},\ and\ \citenamefont {Gauger}}]{sow2019}%
  \BibitemOpen
  \bibfield  {author} {\bibinfo {author} {\bibfnamefont {J.~K.}\ \bibnamefont
  {Sowa}}, \bibinfo {author} {\bibfnamefont {J.~A.}\ \bibnamefont {Mol}}, \
  and\ \bibinfo {author} {\bibfnamefont {E.~M.}\ \bibnamefont {Gauger}},\
  }\href {\doibase 10.1021/acs.jpcc.8b12163} {\bibfield  {journal} {\bibinfo
  {journal} {The Journal of Physical Chemistry C}\ }\textbf {\bibinfo {volume}
  {123}},\ \bibinfo {pages} {4103} (\bibinfo {year} {2019})}\BibitemShut
  {NoStop}%
\end{thebibliography}%

\section*{Appendix}
\appendix
\section{Zero-frequency noise}
\label{App1}
The zero-frequency noise is defined by
\begin{align}
\label{noisesub}
\langle \delta J^2 \rangle &= 2\int_0^\infty dt \langle (J(t)-\langle J \rangle)(J(0)-\langle J \rangle)\rangle \\ \notag &= 2\int_0^\infty dt \langle \delta J(t) \delta J(0) \rangle,
\end{align}
where $\langle J \rangle=t^{-1} \int_0^t dt' J(t')$ denotes the average over a time trajectory.
We now proof the equality (Eq.\ (5) in main text)
\begin{align}
\label{noisesub2}
\langle \delta J^2 \rangle &\equiv 2e^2(\langle n_R^2 \rangle - \langle n_R \rangle^2)/t \\ \notag &=2e^2\langle (n_R(t) -\langle n_R \rangle)(n_R(t) - \langle n_R \rangle) \rangle/t,
\end{align}
where $\langle n_R \rangle= t^{-1}\int_0^t dt' n_R(t')$ is the average particle number and $\langle n_R^2 \rangle= t^{-1}\int_0^t dt' n_R^2(t')$ is the second moment of the particle number in time interval $t$.

To proceed, we define the total number of charges during time $t$ by $n_R(t)-n_R(0)=\int_0^t dt' \dot{n}_R(t')\equiv e^{-1}\int_0^t dt' J(t')$ where we set $n_R(0)=0$. Further we define $\delta n_R(t)=n_R(t)-\langle n_R \rangle=e^{-1}\int_0^t dt' \big (J(t')-\langle J \rangle\big)=e^{-1}\int_0^t dt' \delta J('t)$. The average current during time $t$ reads $\langle J \rangle = e\langle n_R\rangle /t$.
We take the limit of $t\to \infty$ and write the right side of Eq.\ \eqref{noisesub2} as
\begin{align}
\label{noisesub3}
\lim_{t\to\infty}&\frac{2e^2}{t}\langle (n_R(t) -\langle n_R \rangle)(n_R(t) - \langle n_R \rangle) \rangle \\ \notag &=\lim_{t\to\infty} \frac{2}{t} \int_0^t dt' \int_0^t dt'' \langle \delta J(t') \delta J(t'') \rangle
\\ \notag
&=\lim_{t\to\infty} \frac{2}{t} 2\int_0^t dt' \int_0^{t'} dt'' \langle \delta J(t') \delta J(t'') \rangle 
\\ \notag & = \lim_{t\to\infty}  \frac{4}{t}\int_0^t dt' \int_0^{t'} dt'' \langle \delta J(t'-t'') \delta J(0) \rangle
\\ \notag & = \lim_{t\to\infty}  \frac{4}{t}\int_0^t dt' \int_0^{t'} d\tau \langle \delta J(\tau) \delta J(0) \rangle
\\ \notag & = \lim_{t\to\infty}  \frac{4}{t}\int_0^t d\tau \int_\tau^{t} dt' \langle \delta J(\tau) \delta J(0) \rangle
\\ \notag & = \lim_{t\to\infty}  \frac{4}{t}\int_0^t d\tau (t-\tau) \langle \delta J(\tau) \delta J(0) \rangle
\\ \notag & = 4\int_0^\infty d\tau \langle \delta J(\tau) \delta J(0) \rangle= 2\int_{-\infty}^\infty d t \langle \delta J(t) \delta J(0) \rangle\\ \notag
&\equiv \langle \delta J^2 \rangle
\end{align}
We use the fact that one can express the double integration as $\int_0^t dt' (\int_0^{t'} dt'' +\int_{t'}^t dt'')= 2 \int_0^t dt' \int_0^{t'} dt'' $ and the current correlation depends only on time difference $\tau=t''-t'$. We further assume for long times $t\to \infty$ that $\langle \delta J(t) \delta J(0) \rangle\to 0$.
\section{Stationary charge current and zero-frequency noise by counting statistics}
\label{sec2}

In what follows, we use elegant expressions for the velocity and diffusion coefficient obtained by Koza \cite{koz1999} by means of a modified generator.

In order to derive that modified generator we consider the master equation for the occupation and deoccupation process. The probability that the molecule is in an electronic occupied state denotes $P_a$ and in an electronic unoccupied state $P_b$, such that the master equation reads

\begin{align}
\label{master}
\begin{pmatrix}
\dot{P}_a \\
\dot{P}_b
\end{pmatrix} 
= 
\begin{bmatrix}
-(k_{a\to b}^R+k_{a\to b}^L) & (k_{b\to a}^R+k_{b\to a}^L) \\
(k_{a\to b}^R+k_{a\to b}^L) & -(k_{b\to a}^R+k_{b\to a}^L)
\end{bmatrix}
\begin{pmatrix}
P_a \\
P_b
\end{pmatrix}.
\end{align}
To proceed one considers the master equation for $P_a$ and $P_b$ as diffusion along an infinite chain where only nearest neighbor hopping is allowed which represents a transition from molecular state $a \to b$ or $b \to a$.    
In this picture, the master equation can be reformulated for the molecular state $l=a,b$ as
\begin{align}
\label{master3}
\dot{P}&_l(n_R,n_L,t) = \\ \notag &\sum_{K=R;L} \sum_{j=-1}^{+1}[k^K_{l+j,l}P_{l+j}(n_R+j\delta_{KR},n_L+j\delta_{KL},t) \\ \notag &-k^K_{l,l+j}P_l(n_R,n_L,t)].
\end{align}
In this formulation the molecular state changes when $l\to l\pm 1$ while $n_K$ represents a counting index that increases (decreases) by $1$ each time the electron moves to right (left) where we associate the hops on the lattice with electron exchange with the $K=R$ $(L)$ right (left) electrode. $\delta_{KR}(\delta_{KL})=1$ for $K=R(L)$ in Eq.\ (\ref{master3}) is the Kronecker delta. These electron exchanges are described by the following rules:
\begin{align}
a(n_R,n_L)\to b(n_R+1,n_L)&: \hspace*{0.5cm} \text{electron given to R,}\\ \notag
a(n_R,n_L)\to b(n_R,n_L-1)&: \hspace*{0.5cm} \text{electron given to L,}\\ \notag
b(n_R,n_L)\to a(n_R-1,n_L)&: \hspace*{0.5cm} \text{electron taken from R,}\\ \notag
b(n_R,n_L)\to a(n_R,n_L+1)&: \hspace*{0.5cm} \text{electron taken from L},
\end{align}
where $a(n_K)$ ($b(n_K)$) corresponds to molecule in state $a$ ($b$) while the counting index is $n_K$. In this formalism all other transition processes, e.g., $a(n_R,n_L) \to b(n_{R}-1,n_L)$, are forbidden.

We can now calculate the Fourier transform of Eq.\ (\ref{master3})
\begin{align}
\label{master4}
\dot{P}_l(w_R,w_L,t)&=\sum_{K=R;L}\sum_{j=-1}^{+1}[k^K_{l+j,l}e^{jw_K} P_{l+j}(w_R,w_L,t)\\ \notag &-k^K_{l,l+j}P_l(w_R,w_L,t)],
\end{align}
where the Fourier transform is defined here with $P_l(w_R,w_L,t)=\sum_{n_R} \sum_{n_L} \exp{(w_R n_R+w_L n_L)}\\ \times P_l(n_R,n_L,t)$. 
Eq.\ (\ref{master4}) can be written in a compact form using a $L \times L$ matrix, with elements $\Lambda^K_{lj}(k)$, referred as modified generator
\begin{align}
\label{master5}
\dot{P}^K_l(w_R,w_L,t)=\sum_K\sum_{j=-1}^{+1} \Lambda^K_{lj}(w_K)P_j(w_R,w_L,t).
\end{align}

The modified generator $\Lambda(w_R,w_L)=\Lambda^R(w_R)+\Lambda^L(w_L)$ reads
\begin{align}
\label{gen1}
\Lambda&(w_R,w_L)
\\ \notag &= 
\begin{bmatrix}
-k_{a\to b}^R-k_{a\to b}^L & k_{b\to a}^R e^{w_R}+k_{b\to a}^Le^{-w_L} \\
k_{a\to b}^R e^{-w_R}+k_{a\to b}^Le^{w_L} & -k_{b\to a}^R-k_{b\to a}^L
\end{bmatrix}.
\end{align}

The matrix elements $\Lambda^K_{lj}(w^K)$ will be eventually used to determine the charge current and diffusion constant.  Koza proved in a general way for an arbitrary modified generator $\Lambda(w)$ that the matrix $\Lambda(0)$ is irreducible with one eigenvalue $\lambda_0(w=0)=0$ whose correspondent eigenvector has only positive entries \cite{koz1999}. This will be related to the steady state. All other eigenvalues (real part) are negative. The steady state has the form $P(w,t)=\exp{(\lambda_0(w)t)}$ while in our metal-molecular-metal model with two sides $P(w_R,w_L,t)=\exp{(\lambda_0(w_R,w_L)t)}$.

The charge current, say to the right side, is then defined as 
\begin{align}
\label{current}
J&=\lim_{t\to \infty} e \frac{\langle n_R \rangle}{t}=\lim_{t\to \infty} e \frac{\partial_{w_{R}} P(w_R,w_L,t)|_{w_R=0,w_L=0}}{t} \\ \notag &=e \partial_{w_R} \lambda(w_R,w_L)|_{w_R=0,w_L=0},
\end{align}
with elementary charge $e$, and the current noise at zero-frequency reads
\begin{align}
\label{diffusion}
\langle \delta J^2 \rangle&=\lim_{t\to \infty} 2e^2 \frac{\langle n_R^2 \rangle -\langle n_R \rangle^2}{2t}\\ \notag &=\lim_{t\to \infty} 2e^2 \frac{\partial^2_{w_R} P(w_R,w_L,t)|_{w_R=0,w_L=0}}{t} \\ \notag &- \frac{(\partial_{w_R} P(w_R,w_L,t)|_{w_R=0,w_L=0})^2}{t}\\ \notag &=2e^2\partial^2_{w_R} \lambda(w_R,W_L)|_{w_R=0,w_L}.
\end{align}

We can now determine $\langle \delta J_R \rangle$ and $\langle \delta J_R^2 \rangle$ in an elegant way without explicitly calculating a single eigenvalue only by considering the characteristic polynomial.

The characteristic polynomial associated with $\Lambda(w)$ are $\det(\lambda \hat{1}-\Lambda(w_R,w_L))= \sum_n C_n(w_R,w_L)\lambda^n =0$. Since $\lambda_0(w_R,w_L)$ is a root of this characteristic polynomial also the following relation holds 
$\sum_n C_n(w_R,w_L)\lambda_0(w_R,w_L)^n =0$.

Taking the derivative with $w_{R(L)}$ (now portrayed with a prime) and setting $w_{R(L)}=0$ leads to the following expression for the current and diffusion coefficient
\begin{align}
\langle J_R \rangle=e \lambda'=-e\frac{C'_0}{C_1}
\end{align}
and 
\begin{align}
\langle \delta J_R^2 \rangle=2e^2\lambda''=-2e^2\frac{C''_0+2C'_1\lambda'+2C_2(\lambda')^2}{C_1}.
\end{align}
Using characteristic polynomial 

\noindent $[\lambda+(k_{a\to b}^R+k_{a\to b}^L)][\lambda+(k_{b\to a}^R+k_{b\to a}^L)]-[k_{b\to a}^R e^{w_R}+k_{b\to a}^Le^{-w_L}][k_{a\to b}^R e^{-w_R}+k_{a\to b}^Le^{w_L}]=0$ resulting from the modified generator $\Lambda(w_R,w_L)$ of Eq.\ \eqref{gen1} we can determine the current and current noise noting that $\langle J_R \rangle=e \partial_{w_R} \lambda = \langle J_L \rangle=e \partial_{w_L} \lambda=\langle J\rangle$ and 
$\langle \delta J_R^2 \rangle=2e^2 \partial^2_{w_R} \lambda = \langle \delta J_L^2 \rangle=2e^2 \partial_{w_L}^2 \lambda=\langle \delta J^2\rangle$ are equal on both sides.

We find the coefficients
\begin{align}
C_0'&=k_{b\to a}^R k_{a\to b}^L-k_{a\to b}^R k_{b\to a}^L \\ 
C_1&=k_{b\to a}^R + k_{a\to b}^L+k_{a\to b}^R+ k_{b\to a}^L
\\
C_0''&=k_{b\to a}^R k_{a\to b}^L+k_{a\to b}^R k_{b\to a}^L \\
C_1'&=0 \\
C_2&=1.
\end{align}

\section{Entropy production rate}
\label{sec3}

The total change of systems entropy production $\dot{S}(t)$ can be derived from the Boltzmann-Gibbs expression \cite{bol1877,GibbsBook} for the system entropy $S(t)=-k_B\sum_iP_i(t)\ln P_i(t)$ and the master equation of Eq.\ (2) in the main text \cite{sna1976}
\begin{align}
\label{mast2}
\frac{d}{dt}P_i(t)=\sum_j k_{ji}P_j(t),
\end{align}
where the transition rates $k_{ji}$ given state $i$ satisfy  
\begin{align}
\label{cont}
\sum_i k_{ji} =0.
\end{align}
Using these equations Eqs.\ (\ref{mast2}) and (\ref{cont}) the entropy change in the system reads
\begin{align}
\label{Prod}
\dot{S}(t)&=-k_B\sum_i \dot{P}_i \ln P_i \\ \notag
&=-k_B\sum_{ij} k_{ji} P_j \ln P_i = -k_B\sum_{ij} k_{ji} P_j \ln{\frac{P_i}{P_j}} \\ \notag
& = \frac{k_B}{2}\sum_{ij} \bigg( k_{ji} P_j - k_{ij} P_i \bigg) \ln{\frac{P_j}{P_i}} \\ \notag
& = \frac{k_B}{2}\sum_{ij} \bigg( k_{ji} P_j - k_{ij} P_i \bigg) \ln{\frac{k_{ji}P_j}{k_{ij}P_i}} \\ \notag &+ \frac{k_B}{2}\sum_{ij} \bigg( k_{ji} P_j - k_{ij} P_i \bigg) \ln{\frac{k_{ij}}{k_{ji}}} \\ \notag
&= \dot{\sigma}(t)+\dot{S}_e(t).
\end{align}

The first term in the last line of Eq.\ (\ref{Prod}) is the entropy production rate $\dot{\sigma}(t)$ which is always positive due to $(a-b)\ln(a/b)>0$ and the second term is the entropy flow $\dot{S}_e(t)$ into the bath.

In the non-equilibrium steady state $S(t)$ is constant, $\dot{S}(t)=0$, and all entropy generated is continuously been given into the bath, $\dot{S}_e(t)=-\dot{\sigma}(t)$ which leads to Eq.\ (24) in the main text. Note that in most cases the bath is idealized as being without internal dissipation, so that its entropy change only results from the entropy flow from the system. In the equilibrium state, the detailed balance relation $k_{ji}P_j=k_{ij}P_i$ is satisfied such that $\dot{S}_e(t)=\dot{\sigma}(t)=0$ in Eq.\ (\ref{Prod}) and no entropy is produced.

\section{Ratios of rate}
\label{sec4}

One may now transform the integral equations (19) and (20) of the main text with $\epsilon \to \epsilon+e\Delta \Phi_K$ and establish their ratio
\begin{align}
&\frac{k_{a\to b}^K}{k_{b\to a}^K}=\frac{\int d\epsilon \Gamma_K(\epsilon+e\Delta \Phi_K)\frac{\exp[\beta_K\epsilon]}{\exp[\beta_K\epsilon]+1}}{\int d\epsilon \Gamma_K(\epsilon+e\Delta \Phi_K)\frac{1}{\exp[\beta_K\epsilon]+1}}\\ \notag& \times\frac{\exp\bigg[-\frac{\beta_s}{4E_r}(\epsilon+e\Delta \Phi_K+E_r-\epsilon_d)^2\bigg]}{\exp\bigg[-\frac{\beta_s}{4E_r}(\epsilon_d+E_r-\epsilon-e\Delta \Phi_K)^2\bigg]} \\
&=\frac{\int d\epsilon \Gamma_K(\epsilon+e\Delta \Phi_K)\frac{\exp[\beta_K\epsilon]\exp[-\beta_s(\epsilon+e\Delta \Phi_K-\epsilon_d)]}{\exp[\beta_K\epsilon]+1}}{\int d\epsilon \Gamma_K(\epsilon+e\Delta \Phi_K)\frac{1}{\exp[\beta_K\epsilon]+1}}\\ \notag &\times\frac{\exp\bigg[-\frac{\beta_s}{4E_r}(-\epsilon-e\Delta \Phi_K+E_r+\epsilon_d)^2\bigg]}{\exp\bigg[-\frac{\beta_s}{4E_r}(\epsilon_d+E_r-\epsilon-e\Delta \Phi_K)^2\bigg]}.
\end{align}
If $\beta_s\equiv\beta_K\equiv\beta=(k_BT)^{-1}$, their ratio reads
\begin{align}
\label{1}
\frac{k_{a\to b}^K}{k_{b\to a}^K}=\exp\big[-\beta(e\Delta \Phi_K-\epsilon_d)\big].
\end{align}

For later purpose and for $\Gamma(\epsilon) \equiv \Gamma$, $\epsilon_d=0$ and $\Delta \Phi_R=-\Delta \Phi_L=\Delta \Phi/2$ the ratio of the rates $\frac{k_{a\to b}^R}{k_{a\to b}^L}$ can be written as
\begin{align}
&\frac{k_{b\to a}^L}{k_{b\to a}^R}=\frac{\int_{-\infty}^{\infty} d\epsilon\frac{1}{\exp[\beta\epsilon]+1}\exp\bigg[-\frac{\beta}{4E_r}(-\epsilon+e\Delta \Phi/2+E_r)^2\bigg]}{\int_{-\infty}^{\infty} d\epsilon\frac{1}{\exp[\beta\epsilon]+1}\exp\bigg[-\frac{\beta}{4E_r}(-\epsilon-e\Delta \Phi/2+E_r)^2\bigg]} \\
&=\frac{\int_{-\infty}^{\infty} d\epsilon\frac{1}{\exp[\beta\epsilon]+1}\exp\bigg[-\frac{\beta}{4E_r}(\epsilon-e\Delta \Phi/2-E_r)^2\bigg]}{\int_{-\infty}^{\infty} d\epsilon\frac{1}{\exp[\beta\epsilon]+1}\exp\bigg[-\frac{\beta}{4E_r}(-\epsilon-e\Delta \Phi/2+E_r)^2\bigg]}\\
&=\frac{\int_{-\infty}^{\infty} d\epsilon\frac{1}{\exp[-\beta\epsilon]+1}\exp\bigg[-\frac{\beta}{4E_r}(-\epsilon-e\Delta \Phi/2-E_r)^2\bigg]}{\int_{-\infty}^{\infty} d\epsilon\frac{1}{\exp[\beta\epsilon]+1}\exp\bigg[-\frac{\beta}{4E_r}(-\epsilon-e\Delta \Phi/2+E_r)^2\bigg]}
\\& \label{2} =\frac{\int_{-\infty}^{\infty} d\epsilon\frac{\exp[-\beta(\epsilon+e\Delta \Phi/2)]}{\exp[-\beta\epsilon]+1}\exp\bigg[-\frac{\beta}{4E_r}(-\epsilon-e\Delta \Phi/2+E_r)^2\bigg]}{\int_{-\infty}^{\infty} d\epsilon\frac{1}{\exp[\beta\epsilon]+1}\exp\bigg[-\frac{\beta}{4E_r}(-\epsilon-e\Delta \Phi/2+E_r)^2\bigg]} \\ &=\exp\bigg[-\beta e \Delta \Phi/2 \bigg].
\label{2b}
\end{align}
For $\Gamma(\epsilon) \equiv \Gamma$ and $\epsilon_d=0$, the ratio of the rates $\frac{k_{a\to b}^L}{k_{b\to a}^R}$ can be written as
\begin{align}
&\frac{k_{a\to b}^L}{k_{b\to a}^R} \\ \notag &=\frac{\int_{-\infty}^{\infty} d\epsilon\frac{\exp[\beta\epsilon]}{\exp[\beta\epsilon]+1}\exp\bigg[-\frac{\beta}{4E_r}(\epsilon+e\Delta \Phi/2+E_r)^2\bigg]}{\int_{-\infty}^{\infty} d\epsilon\frac{1}{\exp[\beta\epsilon]+1}\exp\bigg[-\frac{\beta}{4E_r}(-\epsilon+e\Delta \Phi/2+E_r)^2\bigg]} \\
&\label{3}=\frac{\int_{-\infty}^{\infty} d\epsilon\frac{\exp[\beta\epsilon]}{\exp[\beta\epsilon]+1}\exp\bigg[-\frac{\beta}{4E_r}(\epsilon+e\Delta \Phi/2+E_r)^2\bigg]}{\int_{-\infty}^{\infty} d\epsilon\frac{\exp[\beta\epsilon]}{\exp[\beta\epsilon]+1}\exp\bigg[-\frac{\beta}{4E_r}(\epsilon+e\Delta \Phi/2+E_r)^2\bigg]}=1
\end{align}

\section{Bound on $Q_2$}
\label{sec5}

In order to proof that $Q=Q_1-Q_2$ is minimal for $\epsilon_d=0$ for symmetric contacts $\Gamma_R=\Gamma_L=\Gamma$ and homogeneous temperatures of leads and environments $\beta_K=\beta_S=\beta$ we realize that $Q_1=e\Delta \Phi \coth[\beta e\Delta \Phi/2]$ for all junction parameters (see discussion in main text). 
For $Q=Q_1-Q_2$ minimal we need to proof that $Q_2$ is maximal for $\epsilon_d=0$.

Using relation in Eq.\ \eqref{1} we can write for $\epsilon_d \neq 0$
\begin{widetext}
\begin{align}
Q_2&=2e\Delta \Phi\frac{k_{b\to a}^R k_{a\to b}^L-k_{a\to b}^R k_{b\to a}^L}{(k_{b\to a}^R + k_{a\to b}^L+k_{a\to b}^R+ k_{b\to a}^L)^2} \\ 
\label{33}
&=2e\Delta \Phi \frac{k_{a\to b}^L}{k_{b\to a}^R }\frac{1-\exp{[-\beta e\Delta \Phi]}}{\bigg(1+\exp{[-\beta (e\Delta \Phi/2-\epsilon_d)]}+\frac{k_{a\to b}^L}{k_{b\to a}^R }\big[1+\exp{[\beta (-e\Delta \Phi/2-\epsilon_d)]}\big]\bigg)^2}.
\end{align}
\end{widetext}
For $\epsilon_d=0$ and $\frac{k_{a\to b}^L}{k_{b\to a}^R }=1$, see relation \eqref{3}, such that
\begin{align}
\label{34}
Q_2^{\epsilon_d=0}&=2 e\Delta \Phi \frac{1-\exp{[-\beta e\Delta \Phi]}}{\big(2+2\exp{[-\beta e\Delta \Phi/2]}\big)^2}
\\ \notag &=\frac{e\Delta \Phi}{2} \tanh{[-\beta e\Delta \Phi/4]}.
\end{align} 
We now want to show that Eq.\ \eqref{34} is larger or equal than Eq.\ \eqref{33} which leads to the inequality $Q_2^{\epsilon_d=0}\geq Q_2$ which can be written as
\begin{widetext}
\begin{align}
\label{35}
\frac{k_{a\to b}^L}{k_{b\to a}^R } \frac{\big(2+2\exp{[-\beta e\Delta \Phi/2]}\big)^2}{\bigg(1+\exp{[-\beta (e\Delta \Phi/2-\epsilon_d)]}+\frac{k_{a\to b}^L}{k_{b\to a}^R }\big[1+\exp{[\beta (-e\Delta \Phi/2-\epsilon_d)]}\big]\bigg)^2} \leq 1.
\end{align}
\end{widetext}

We write the ratio $\frac{k_{a\to b}^L}{k_{b\to a}^R}$ for $\epsilon_d\neq 0$ as
\begin{align}
\label{36}
a&=\frac{k_{a\to b}^L}{k_{b\to a}^R} \\ \notag &=\frac{\int_{-\infty}^{\infty} d\epsilon\frac{\exp[\beta\epsilon]}{\exp[\beta\epsilon]+1}\exp\bigg[-\frac{\beta}{4E_r}(\epsilon+e\Delta \Phi/2+E_r-\epsilon_d)^2\bigg]}{\int_{-\infty}^{\infty} d\epsilon\frac{\exp[\beta\epsilon]}{\exp[\beta\epsilon]+1}\exp\bigg[-\frac{\beta}{4E_r}(\epsilon+e\Delta \Phi/2+E_r+\epsilon_d)^2\bigg]},
\end{align}
where $a\geq 0$. The inequality of Eq.\ \eqref{35} then reads
\begin{align}
\label{37}
a&\big(2+2\exp{[-\beta e\Delta \Phi/2]}\big)^2 \\ \notag &\leq \bigg(1+a+\exp{[-\beta e\Delta \Phi/2]}\big[\exp{[\beta\epsilon_d]}+a\exp{[-\beta\epsilon_d]}\big]\bigg)^2.
\end{align}
It is easy to show that the term $\big[\exp{[\beta\epsilon_d]}+a\exp{[-\beta\epsilon_d]}\big]\geq 2\sqrt{a}$ on the right hand side of Eq.\ \eqref{37}. Using this bound, we recast Eq.\ \eqref{36} to 
\begin{align}
\label{38}
a\big(2&+2\exp{[-\beta e\Delta \Phi/2]}\big)^2 \\ \notag &\leq \big(1+a+2\sqrt{a}\exp{[-\beta e\Delta \Phi/2]}\big)^2 \\ 
\label{39}
0 &\leq 1+a^2+2(1+a)\sqrt{a}W-2a(1+2W),
\\ 
\label{40}
0 &\leq (a-1)^2+2\sqrt{a}(1-\sqrt{a})^2W,
\end{align}
with $W=2\exp{[-\beta e\Delta \Phi/2]}$. Since $a\geq 0$ and $W\geq 0$, the inequality in Eq.\ \eqref{40} is always fulfilled and therefore $Q_2^{\epsilon_d=0}\geq Q_2$. Note that $a=1$ is the case when $\epsilon_d=0$ in Eq.\ \eqref{36} and $Q_2^{\epsilon_d=0}= Q_2$ (see analytic form in Eq.\ \eqref{34}).

\section{$g(\Delta \Phi)$ for several reorganization energies}
\label{sec6}
\begin{figure}[h!!!!!]
\centering
\includegraphics[width=\linewidth]{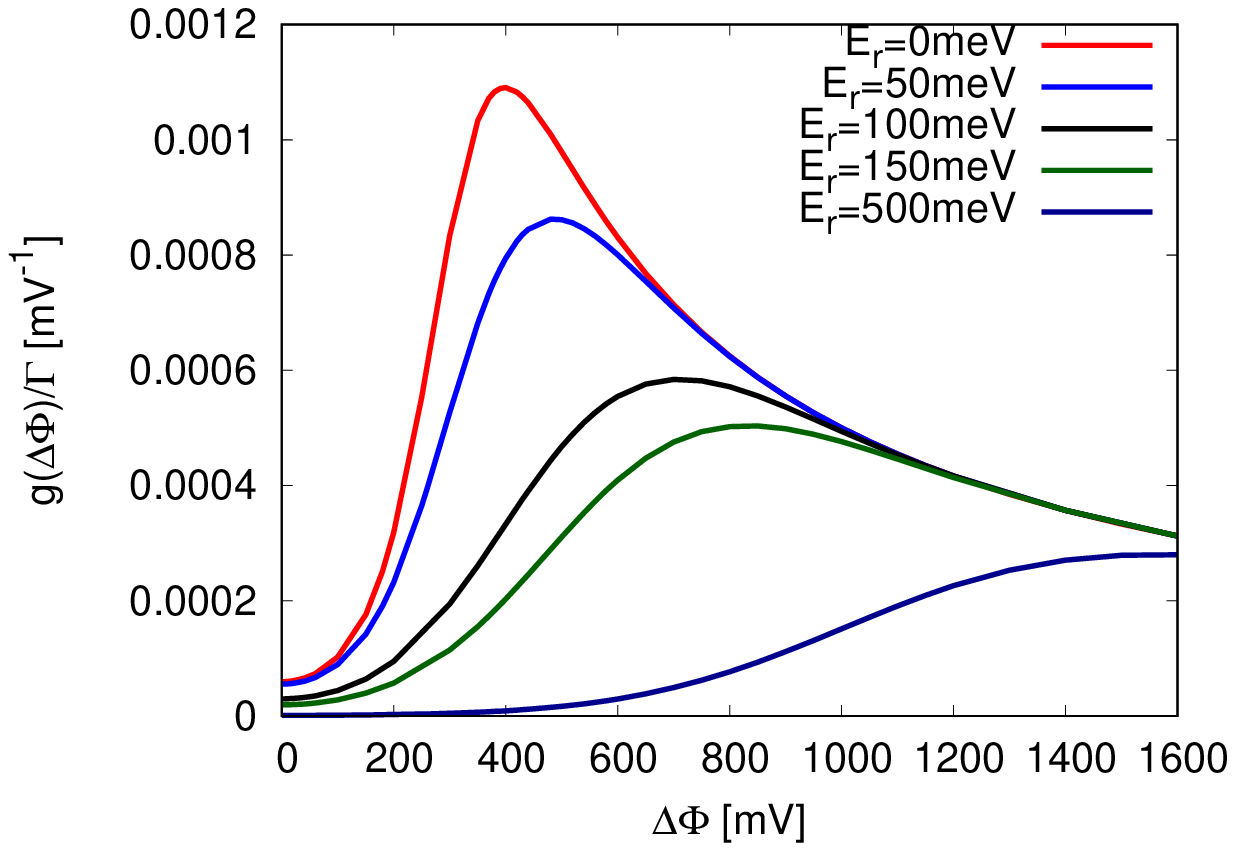}
\caption{\label{figConduct}$g(\Delta \Phi)=\langle J \rangle /\Delta \Phi$ for different $E_R$ (Marcus model) plotted against the applied bias potential $\Delta \Phi$. The temperature is $T=300$K ($k_BT\simeq 26$meV) and $\epsilon_d=150$meV. Note that for $E_R=0$ we set the transfer rates to $k_{a\to b}^K=\Gamma_K [1-f_K(\beta_K,\epsilon_d)]$ and $k_{b\to a}^K=\Gamma_K f_K(\beta_K,\epsilon_d)$.}
\end{figure}

We determine $g(\Delta \Phi)=\langle J \rangle /\Delta \Phi$, Fig.\ \ref{figConduct}, for different reorganization energies $E_R$ (Marcus model) in dependence of different applied bias potential $\Delta \Phi$.

\section{Current and current noise in a Photovoltaic cell model}
\label{sec7}

By following the same procedure as in Section \ref{sec2}, we first determine the modified generator for the master equation (32) in the main text to 

\begin{align}
\Lambda&(w_L,w_R) \\ \notag &=
\begin{bmatrix} 
		-(k_{0L}+k_{0R}) & k_{L0}e^{-w_L} & k_{R0}e^{w_R} \\ 
		k_{0L}e^{w_L} & -(k_{L0}+k_{LR}) & k_{RL} \\ 
		k_{0R}e^{-w_R} & k_{LR} & -(k_{R0}+k_{RL})
	\end{bmatrix}.
\end{align}
The same procedure as in Section 2 leads to the following expression for the current

\begin{align}
\langle J \rangle=e \lambda'=-e\frac{C'_0}{C_1}
\end{align}
and  zero-frequency noise

\begin{align}
\langle \delta J^2 \rangle=2e^2\lambda''=-2e^2\frac{C''_0+2C'_1\lambda'+2C_2(\lambda')^2}{C_1}.
\end{align}

We find the coefficients
\begin{align}
C_0'&=k_{L0}k_{RL}k_{0R}-k_{0L}k_{LR}k_{R0} \\ 
C_1&=k_{R0}(k_{L0}+k_{LR}+k_{0L})+k_{RL}(k_{0L}+k_{0R}+k_{L0}) \\ \notag &+k_{0L}k_{L0}+k_{LR}(k_{0L}+k_{0R})
\\
C_0''&=-(k_{L0}k_{RL}k_{0R}+k_{0L}k_{LR}k_{R0})\\
C_1'&=0 \\
C_2&=k_{0L}+k_{0R}+k_{L0}+k_{LR}+k_{R0}+k_{RL}.
\end{align}

\section{Bound on the TUR for photovoltaic cell}
\label{sec8}
\begin{figure}[h!!!!!]
\centering
\includegraphics[width=\linewidth]{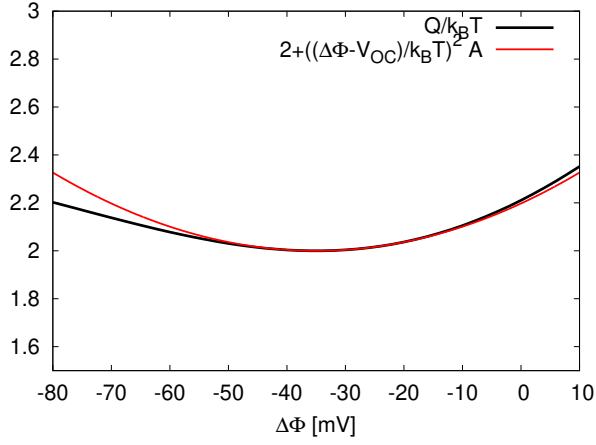}
\caption{\label{fig2S} $Q/k_BT = \dot{\sigma} k_B^{-1}\langle \delta J^2 \rangle/ 2\langle J\rangle ^2$ (black line) and the approximation of Eq.\ \eqref{expand2} (red line) against the applied bias potential $\Delta \Phi$. The temperature is $T=300$K and $k_BT_K= 26$meV, the "sun" temperature $T_S=461$K and $k_BT_S=40$meV. The energy difference of the upper and lower level is chosen to be $\Delta E=100$meV=$100\hbar \Gamma$ where $\Gamma$ is the electron transfer rate to each lead while we set $\hbar \equiv 1$. We choose the transfer rate between the two energy levels to $\Gamma_S=\Gamma$. The open circuit voltage is $V_{oc}=-(1-T/T_S)\Delta E/e=-35$mV.}
\end{figure}

We determine the bound on the TUR for the photovoltaic cell. The TUR product reads
\begin{align}
\label{TURphotocell}
 Q/k_BT &= \dot{\sigma} k_B^{-1}\langle \delta J^2 \rangle/2 \langle J\rangle^2 \\ \notag 
  &= Q_1-Q_2
 \\ \label{TURphoto} &= k_B^{-1}e\bigg [T^{-1}[-\Delta \Phi - \Delta E/e]+T_S^{-1}\Delta E/e \bigg]  \times \\ \notag &\bigg [ \frac{(k_{L0}k_{RL}k_{0R}+k_{0L}k_{LR}k_{R0})}{(k_{L0}k_{RL}k_{0R}-k_{0L}k_{LR}k_{R0})} \\ 
 \label{TURphotob} &- \frac{(k_{0L}+k_{0R}+k_{L0}+k_{LR}+k_{R0}+k_{RL})}{C_1^2} \times  \\ \notag &2(k_{L0}k_{RL}k_{0R}-k_{0L}k_{LR}k_{R0})\bigg].
\end{align}
For $\Delta \Phi \to V_{oc}=-(1-T/T_S)\Delta E /e$, the second term $Q_2$ (Eq.\ \eqref{TURphotob}) vanishes because it is proportional to $1-\exp{[\beta (e\Delta \Phi +\Delta E)-\beta_S\Delta E]}$ which for $\lim \Delta \Phi \to V_{oc}$ is zero. 
Define $C=k_B^{-1}e[-T^{-1}[-\Delta \Phi - \Delta E/e]-T_S^{-1}\Delta E/e ]$ we can write the remaining $Q_1$ in Eq.\ \eqref{TURphoto} as
\begin{align}
\label{TURphoto2}
 Q/k_BT &= -C \bigg [ \frac{1+\exp{C}}{1-\exp{C}}\bigg]=C \coth[C/2] \\ \notag &= C \bigg[\frac{2}{C}+ \sum_{k=1}^{\infty} \frac{C}{k^2 \pi^2 +(C/2)^2}\bigg] \geq 2,
\end{align}
which fulfills the inequality. 
For $\Delta \Phi \to V_{oc}$ we easily see $C\to 0$ that Eq.\ \eqref{TURphoto2} reads
\begin{align}
 Q/k_BT = C \bigg [ \frac{2+C}{C}\bigg]=2.
\end{align}
Consider $Q/k_BT$ in dependence of the coefficient $C$ such that Eq.\ \eqref{TURphotocell} reads
\begin{align}
 &Q/k_BT = \label{expand} C \bigg[\frac{2}{C}+ \sum_{k=1}^{\infty} \frac{C}{k^2 \pi^2 +(C/2)^2} \\ \notag
 &+ \frac{(k_{0L}+k_{0R}+k_{L0}+k_{LR}+k_{R0}+k_{RL})}{C_1^2} \times \\ \notag & 2k_{L0}k_{RL}k_{0R}(1-\exp{C})\bigg].
\end{align}
Near the stopping voltage $C\to 0$ we can expand Eq.\ \eqref{expand} up to order $\mathcal{O}[C^2]$ such that 
Eq.\ \eqref{expand} reads
\begin{align}
Q/k_BT &\simeq \label{expand2} 2 + C^2 \bigg [ \frac{1}{6}- \frac{(2\Gamma+\Gamma_S(1+2n(\beta_S\Delta E))}{C_1(V_{OC})^2} \times \\ \notag &2 k_{L0}(V_{OC})k_{0R}(V_{OC})\Gamma_S(1+n(\beta_S\Delta E))\bigg]\\ \notag
&=2+C^2\cdot A=2+[(\Delta\Phi -V_{oc})/k_BT]^2\cdot A,
\end{align}
with the Bose-Einstein distribution $n(\beta_S\Delta E)$ and the transfer rates (Eqs. (35) and (36) in main text) evaluated at $V_{OC}$ where $A>0$. Fig.\ \ref{fig2S} portrays the case of equal transfer rates $\Gamma=\Gamma_S$ and clearly shows that the quadratic term in $C$ dominates near $V_{OC}$. Away from $V_{OC}$ the TUR product $Q$ becomes proportional to the applied bias potential $\Delta \Phi$ such that the minimal value of $Q/k_BT=2$ is reached at $V_{OC}$.

\end{document}